\documentclass[sigconf,nonacm]{acmart}
\usepackage{amsthm}
\usepackage{graphicx}  
\usepackage{subcaption}
\usepackage{tabularx}
\usepackage{colortbl} 
\usepackage{siunitx}
\usepackage{booktabs} 
\usepackage{caption} 
\usepackage{amsmath}
\usepackage{makecell}
\usepackage{array}
\usepackage{xcolor}
\usepackage{listings}
\usepackage{multirow} 
\usepackage{amsmath}

\definecolor{myGreen}{RGB}{255, 255, 255}
\definecolor{myRed}{RGB}{255, 255, 255}
\definecolor{myLightGreen}{RGB}{255, 255, 255}
\definecolor{myLightRed}{RGB}{255, 255, 255}
\definecolor{dkgreen}{rgb}{0,0.6,0}

\begin{document}
\title{100x Cost \& Latency Reduction: Performance Analysis of AI Query Approximation using Lightweight Proxy Models}

\author{Yeounoh Chung, Rushabh Desai, Jian He, Yu Xiao, Thibaud Hottelier, \\Yves-Laurent Kom Samo, Pushkar Khadilkar, Xianshun Chen, Sam Idicula, \\Fatma \"{O}zcan, Alon Halevy, Yannis Papakonstantinou}
\thanks{This paper has been accepted for publication at the 2026 ACM SIGMOD International Conference on Management of Data (SIGMOD '26)}
\affiliation{%
  \institution{Google Cloud}
  \city{Sunnyvale}
  \country{USA}
}
\email{{yeounoh, rsude, hejia, xiayuo, tbh, komsamo, kpushkar, xianshun, samidicula, fozcan, halevy, yannispap}@google.com}

\renewcommand{\shortauthors}{Chung et al.}
\newcommand{\gemini}{\textit{gemini-2.5}~}
\newcommand{\yeounoh}[1]{#1}
\newcommand{\yannis}[1]{\textcolor{blue}{YP: #1}}
\newcommand{\fatma}[1]{\textcolor{pink}{FO: #1}}
\newcommand{\tbh}[1]{\textcolor{teal}{TH: #1}}
\newcommand{\yl}[1]{\textcolor{blue}{YL: #1}}
\newcommand{\alon}[1]{\textcolor{teal}{AH: #1}}
\newcommand{\sam}[1]{\textcolor{blue}{SI: #1}}

\begin{abstract}
Several data warehouse and database providers have recently introduced extensions to SQL called AI Queries,  enabling users to specify functions and conditions in SQL that are evaluated by LLMs, thereby broadening significantly the kinds of queries one can express over the combination of structured and unstructured data. LLMs offer remarkable semantic reasoning capabilities, making them an essential tool for complex and nuanced queries that blend structured and unstructured data. While extremely powerful, these AI queries can become prohibitively costly when invoked thousands of times. 
 
This paper provides an extensive evaluation of a recent AI query approximation approach that enables low cost analytics and database applications to benefit from AI queries.
The approach delivers >100x cost and latency reduction for the semantic filter (AI.IF) operator and also important gains for semantic ranking (AI.RANK). The cost and performance gains come from utilizing cheap and accurate proxy models over embedding vectors. 
We show that despite the massive gains in latency and cost, these proxy models preserve accuracy and occasionally improve accuracy across various benchmark datasets, including the extended Amazon reviews benchmark that has 10M rows.
We present an OLAP-friendly architecture within Google \textit{BigQuery} for this approach for purely online (ad hoc) queries, and a low-latency HTAP database-friendly architecture in \textit{AlloyDB} that could further improve the latency by moving the proxy model training offline.
We present techniques that accelerate the proxy model training. 
\end{abstract}


\maketitle

\section{Introduction}
\textbf{The Cost and Latency Challenges of Augmenting SQL with AI Operators.}
Large Language Models (LLMs) are transforming data analytics by enabling new ways to query both structured and unstructured data, extracting value from unstructured information, and automating semantic insights within data platforms. 
\yeounoh{This transformation is evident across both academic research~\cite{jo2024thalamusdb,patel2024lotus,russo2025abacus,shankar2024docetl} and major cloud data platforms (e.g., Google BigQuery, AlloyDB, Databricks, and Snowflake \cite{google_bqml,databricks_ai_functions,snowflake_aisql,google_alloydb_ai}),} all of which are integrating LLMs to implement AI semantic operators, like semantic search, content summarization and intent classification.
Example~\ref{ex:simple_query} shows a query with a semantic filter on retail reviews.

\begin{example}
\label{ex:simple_query}
A query that filters out negative reviews:
\begin{lstlisting}[
  language=SQL,
  showspaces=false,
  numberstyle=\tiny,
  basicstyle=\ttfamily,
]
SELECT review
FROM amazon_polarity.reviews
WHERE AI.IF("The review is positive: ", review);
\end{lstlisting}
\end{example}

\yeounoh{However, integrating these powerful LLMs but resource-intensive LLM operators into database systems presents significant challenges regarding cost and latency. In large-scale Online Analytical Processing (OLAP) databases, the computational cost makes the LLM-based semantic processing economically unfeasible on millions of rows. In Hybrid Transactional Analytical Processing (HTAP) databases, the latency incurred by LLM calls severely restricts practical applications.
Existing works on LLM-integration techniques often assume small tables or fail to adequately address the performance and cost implications of applying semantic operations.}

\noindent\textbf{Lightweight Proxy Model for Fast and Cost-Efficient AI Query Engine}
\yeounoh{Prior optimization for AI and semantic query processing focused on predictive row skipping to reduce the number of LLM calls for approximate query processing~\cite{jo2024thalamusdb,patel2024lotus} and more execution plan-level optimization techniques, like operator re-ordering or model cascading~\cite{russo2025abacus,patel2024lotus,shankar2024docetl}.
Crucially, these efforts often dismissed simpler non-LLM proxy model-based approximation, suggested by UQE~\cite{dai2024uqe}, as underperforming or insufficiently accurate compared to their more advanced techniques~\cite{patel2024lotus}. 
We argue that this proxy approach deserves a careful revisit. Our core insight is that meeting the aggressive performance requirements of large-scale OLAP and ultra-fast HTAP necessitates a more cost-aggresive strategy -- one that reserves costly LLM invocations for specific, necessary tasks, allowing the lightweight, non-LLM based proxy to handle the majority of queries with high accuracy. }

Following the UQE~\cite{dai2024uqe} strategy, we evaluate lightweight ``proxy'' models as first-class citizens to deliver a high-quality, cost-effective and ultra-fast AI query engine at scale. 
Namely, we utilize cheap and accurate ML models over embedding vectors as ``proxy'' to LLMs. We show that this simple, disregarded approach can be very effective in handling simple semantic queries that can be reduced to  classification tasks.
Such proxy (classification) models , when used appropriately and in conjunction with LLMs, can dramatically reduce the cost and improve the performance of AI operators for complex semantic data analytics.
Unlike existing AI query engines and operator optimization frameworks that prioritize accuracy~\cite{shankar2024docetl} or minor cost reductions through algorithmic optimization and model cascading~\cite{patel2024lotus,russo2025abacus}, our aggressive objective is to minimize unnecessary LLM dependency while preserving quality -- and use LLM only if its powerful reasoning capabilities are essential.

\begin{table}[t!]
\centering
\caption{Latency and cost gains of lightweight proxy models on the query in Example~\ref{ex:simple_query} using a 10-million row table.}
\label{tab:intro_highlight}
\resizebox{\columnwidth}{!}{
\begin{tabular}{l|c|c}
\toprule
\textbf{Approach} & \textbf{Latency Improvement} & \textbf{Cost Savings} \\
\midrule
Online Proxy Model & 329x & 728x \\
Offline Proxy Model & 991x & 792x \\
\bottomrule
\end{tabular}
}
\end{table}

Table~\ref{tab:intro_highlight} illustrates the dramatic latency and cost gains measured against the \textit{BigQuery} (OLAP)
AI query engine with distributed execution for LLM-based AI operators. 
The proxy model architecture achieves 329x latency improvement and 728x cost savings when deployed with proxy models trained online for ad hoc queries, sifting through 10-million row table for semantic filter (AI.IF) operation.
With offline pre-trained proxy models it achieves 991x latency improvement and 792x cost savings, needed for database applications with sub-second latency requirements.

While proxy model-based approximations have been explored in prior works~\cite{patel2024lotus,dai2024uqe}, they have often fallen short, demonstrating sub-optimal performance against complex benchmarks and tasks. For instance, \cite{patel2024lotus} and similar studies have specifically demonstrated the limitations of AI UDFs and similar proxies~\cite{dai2024uqe}  in maintaining quality when blindly substituting LLMs for all use cases. 
In fact, when applied to the right problem, even simple and cheap  conventional ML models (e.g., embedding-based text classification~\cite{text2vec_vectorization,openai_classification_embeddings,tensorflow_word_embeddings}) can outperform LLM-based AI query semantic operation (Table~\ref{tab:simple_binary_classification}).

\yeounoh{We are revisiting this strategy, which leverages a non-LLM based lightweight proxy model to circumvent the inherent cost and latency issues of LLM-based AI operators. This approach is critical for meeting the performance demands of Google \textit{BigQuery} (large-scale OLAP) and \textit{AlloyDB} (high-throughput, low-latency HTAP database) workloads.}


\noindent\textbf{Contributions and Outline.} As the main contribution of this paper, we present and evaluate  low-latency and cost-efficient AI query engine leveraging lightweight (non-LLM) proxy model approximation.
Our approach is built on the observation that many high-frequency semantic operations, such as semantic filter and rank operators, can be reformulated as simple classification problems.
We implemented and evaluated a prototype AI query engine for semantic filter and rank operators in Google \textit{BigQuery} and \textit{AlloyDB} 
to demonstrate massive latency and cost gains while preserving accuracy over various benchmark datasets, including one with 10-million rows.
To the best of our knowledge, this is the first work thoroughly evaluating the implications and potential of proxy models for AI query analytics in million-row table production settings, as well as for low latency database applications.
In Section~\ref{sec:case_for_proxy_model} we present a case for lightweight (classification) proxy model for efficient AI query engine with a formal definition of ``proxy'' model. 
In Section~\ref{sec:proxy_approximation}, we describe the proxy model approximation process. Namely, we describe the online and offline proxy model training strategies, imbalanced proxy model training techniques, the automatic evaluation of proxy models and the adaptive proxy model selection based on the approximation performance. Note, these proper training techniques account for the success of proxy models in our study and likely account for lukewarm performance and quality in prior studies.
In Section~\ref{sec:adaptive_ranker_selection} we further discuss adaptive proxy model selection. This involves moving beyond a simple choice between one proxy model and the LLM baseline to selecting from a set of multiple, more sophisticated proxy models -- not limited to just simple classification models.
We present extensive performance benchmarking results and detailed analysis of the prototype systems for semantic filter and rank operators in Section~\ref{sec:results}. We discuss lessons for future work in Section~\ref{sec:discussions}. We conclude in in Section~\ref{sec:conclusion} with a summary of our key findings and a discussion of future research directions and open challenges inspired by our results.

\section{Related Works}
\label{sec:related_works}

Universal Query Engine (UQE)~\cite{dai2024uqe} demonstrated the efficacy of using embedding-based sample proxy models for semantic filtering, which serves as a key inspiration for our work. UQE also enabled conditional aggregation, which is achieved by employing stratigied sampling methods to select representative samples and then use the LLM to statistically estimate the final aggregate results.
Other contemporary LLM-centric semantic query systems~\cite{jo2024thalamusdb,patel2024lotus,russo2025abacus,shankar2024docetl,palimpzest_cidr2025,Dorbani2025BeyondQ} focus on optimizing the LLM execution plan. Their core strategies are designed to reduce the number of costly LLM calls through techniques like predictive row skipping, operator re-ordering, and model cascading (substituting in smaller LLMs; \cite{patel2024lotus,jo2024thalamusdb,urban2024eleet} also provide accuracy guarantees for the sample or model cascading based  approximation). 
\yeounoh{These works consider the simple non-LLM proxy machine learning model approximation (e.g., ML UDFs and UQE~\cite{dai2024uqe}) as a baseline for comparative study, demonstrating that it lacks the desired accuracy compared to their methods. This skepticism overlooks the unique potential of the proxy approach when applied aggressively and selectively, particularly in scenarios demanding high throughput and low latency.}

In contrast, the core objective of our work is to analyze and demonstrate the feasibility of more aggressive LLM substitution for high-volume semantic operators (semantic filtering and ranking). Instead of focusing on reducing the number of LLM calls or substituting smaller LLMs, our approach introduces a lightweight proxy that directly \yeounoh{replace the LLM invocations -- we fall back to LLM only when the proxy model's approximation falls below a predetermined threshold}. This substitution allows the system to efficiently process the entire table (not just a sample, unlike other systems) and bypasses the computational and cost bottlenecks of LLM inference entirely. Our extensive evaluation is specifically designed to prove that this proxy model approach is a highly viable, cost-efficient, and performant alternative that eliminates the LLM inference cost entirely for core analytic operations.
Our method is designed to be orders of magnitude faster and cheaper by entirely substituting the LLM inference step with a proxy model.

\section{A Case for Lightweight Proxy Model For Efficient AI Query Engine}
\label{sec:case_for_proxy_model}

\begin{table}[th]
\begin{minipage}{\columnwidth}
\centering
\caption{Proxy models (Embedding + Logistic Regression) VS. LLM-based baseline performance and normalized latency improvement against the LLM baseline for AI.IF using Kaggle SPAM email classification dataset.}
\label{tab:simple_binary_classification}
\resizebox{\columnwidth}{!}{
\begin{tabular}{ll|c|c|c}
\toprule
\multicolumn{2}{l|}{\textbf{Metric / Data Size}} & \textbf{Offline Proxy} & \textbf{Online Proxy} & \textbf{LLM} \\ 
\midrule
\multicolumn{1}{l}{Accuracy} & 1115 & 0.99 & 0.98 & 0.96 \\
\midrule
\multirow{2}{*}{\centering Latency $\uparrow$} & 1115 & 634x & 5.4x & 1x \\
& 100K & 632x & 294x & 1x \\ 
\bottomrule
\end{tabular}
}
\end{minipage}
\end{table}

Current AI operator optimization efforts largely concentrate on model cascading or  cost-efficient algorithmic improvements of LLM-based implementations~\cite{palimpzest_cidr2025,VeltriSBP25,shankar2024docetl,patel2024lotus,khattab2023dspy}. 
While these advancements are valuable, they primarily enhance individual semantic operation efficiency and often overlook the potential of non-LLM alternatives. Table~\ref{tab:simple_binary_classification} demonstrates that, in certain scenarios, proxy models can achieve comparable, or even better results at a significantly lower cost (no LLM calls) and latency. 
It is interesting to see that proxy models can sometimes achieve better accuracy results, and we discuss this further in Section~\ref{sec:filtering_results}.

We trained simple proxy models from NLP text classification tutorials~\cite{text2vec_vectorization,openai_classification_embeddings,tensorflow_word_embeddings}, such as text embedding-based Logistic Regression (LR) classifier, with 200 training examples labeled with \gemini.
We compare the proxy models prepared \textit{offline} and \textit{online}, where online means that the sampling, labelling and training of the proxy model are performed within the execution context of the AI query.
Offline proxy models are crucial for database applications with sub-second latency requirements, and the online proxy model training~\cite{dai2024uqe} is invaluable for ad hoc analytical queries.
The offline training uses train data split, and online training happens directly with evaluation with the test data split (1115 email texts).
The proxy models and the LLM baseline both score above 90\% accuracy, and the proxy models yield slightly superior accuracy, 97\%, for this particular dataset~\footnote{\url{https://www.kaggle.com/datasets/jackksoncsie/spam-email-dataset}}.

We also replicate the orginal dataset to construct a larger dataset to highlight the latency gain at scale, up to 100K rows as shown in Table~\ref{tab:simple_binary_classification}.
The proxy models are orders of magnitude faster than the LLM baseline. While the offline latency reduction is massive due to the proxy model being pre-trained, this gain remains static across different table sizes. This is because both the proxy model's prediction time and the LLM's inference time both grow proportionally to the data sizes. 
For the online use case, the latency gain is further amplified by increasing table size, as the overhead of sampling, labeling, and training can be amortized over processing a larger dataset.

We \yeounoh{implement a prototype AI query engine leveraging lightweight proxy models. This framework focuses on applying proxy model-based approximation to dramatically imporve the latency and the cost of AI query and calls LLMs selectively when the quality of approximation falls short.}
We formally define a proxy model as follows:

\begin{definition}
Let M be a complex model (e.g., an LLM, $M_{LLM}$) that maps an input $x \in \mathcal{X}$ to an output $y \in \mathcal{Y}$, such that $y = M(x)$. A proxy model $M_P$ is a simpler model that maps $x \in \mathcal{X}$ to an output $y_P \in \mathcal{Y}$, such that $y_P = M_P(x)$ and $M_P(x) \approx M(x)$  for a relevant subset of $\mathcal{X}$.  The objective of $M_P$ is to approximate the behavior of the complex model, like $M_{LLM}$ for a specific task or query.
\end{definition}


We posit that  proxy models can bring about massive latency and cost gains without compromising the quailty, when used for the right problems. In Section~\ref{sec:results} we present \yeounoh{extensive evaluation} results demonstrating these advantages for semantic filtering and ranking.

\section{Lightweight Proxy Approximation}
\label{sec:proxy_approximation}
The core contribution of our approach is the demonstration that simple, automated techniques are sufficient to efficiently approximate LLMs for high-frequency semantic operators without requiring manual expert intervention. We highlight the following architectural principles:\\
\noindent\textbf{Fully Automated Workflow:} Every incoming semantic operator is treated as a candidate for proxy model training. The system automatically parses each operator into its core components, as depicted in Figure~\ref{fig:ai_query_phys_plan} -- the operator type ($O$), the semantic query or the AI operator prompt ($Q$), and the unstructured column reference ($C$) -- before initiating the proxy model approximation process. All stages of the pipeline, including embedding generation, data sampling, LLM labeling, and model fitting, are fully automated. This process is identical for both online (ad hoc) and offline (pre-trained) settings, ensuring that the training costs for a given table size remain fixed across workloads.\\ 
\noindent\textbf{Elimination of Expert ML Intervention:}
Our methodology explicitly avoids the need for specialized ML expertise or manual hyperparameter tuning. The system utilizes standard "off-the-shelf" models, primarily embedding-based Logistic Regression (LR). 
We utilize a single set of default parameters (\textit{sklearn} default for Logistic Regression, except \textit{class\_weight=``balanced''}) for all datasets and queries. As demonstrated in our ablation study (Section~\ref{sec:learning_tuning}), expensive hyperparameter searches yield negligible benefits over default settings when utilizing high-quality embeddings. 
The system automatically monitors proxy accuracy against LLM labels during training. It employs an ``Adaptive Proxy Model Selection'' mechanism that only deploys the proxy if its performance is within a specific quality threshold ($\tau$) of the LLM baseline; otherwise, it automatically falls back to the LLM baseline.

\subsection{Online and Offline Proxy Model Training}

The training process typically begins with the generation of embeddings \yeounoh{using an embedding model to compute representations for the unstructured data (e.g., text)}, potentially leveraging pre-computed embeddings if available \yeounoh{from previous runs or provided by users}. 
A following sampling step then efficiently selects a small, representative subset of rows from the large table/dataset. Subsequently, the LLM is invoked for labeling on this sampled data, \yeounoh{only} providing ground truth labels for training the proxy model. By limiting LLM calls to a small sample, the overall cost and latency are significantly reduced compared to full-dataset LLM invocation. Beyond classification, embedding vectors also provide a strong baseline for ranking operations; for instance, LOTUS~\cite{patel2024lotus} utilizes embedding-based similarity to facilitate semantic JOIN optimization.

Leveraging proxy models for processing AI queries necessitates effective strategies for their training, which can be broadly categorized into offline and online approaches.

\yeounoh{
\noindent\textbf{Online Training:}
For \textit{BigQuery}, a large-scale OLAP cloud database, we employ online training designed for new incoming semantic queries. The high cost of this online proxy model training is effectively amortized across the massive query processing of millions or billions of rows. The proxy model is dynamically trained during query execution, an approach inspired by the active learning and dynamic techniques used in UQE\cite{dai2024uqe} for semantic filtering.
}

\yeounoh{
\noindent\textbf{Offline Training:}
The primary motivation for offline proxy model training in \textit{AlloyDB}, a high-performance HTAP database is to ensure low latency for AI queries, especially for interactive database applications.
For offline training, proxy models are prepared for known or frequently re-occurring semantic tasks and queries. 
}
By performing the training process offline, we can decouple the overhead of proxy model training entirely from the online query execution path.
\yeounoh{One major advantage of offline training is that you can scale the test-time compute~\cite{Snell2024ScalingLT,muennighoff2025s1} to improve the LLM labeling accuracy (e.g., extra thinking and verification), that are too costly and time consuming for online labeling.}
This decoupled environment also provides the time and resources for systematic model optimization. We can perform more extensive proxy model selection and hyperparameter tuning, to ensure the proxy model is accurate and efficient for known workload patterns before deployment.
While highly effective, the offline-trained proxy models also introduce specific model robustness challenges. In particular, we need to ensure that the trained model remains valid and performs well, even as the data changes. This requires periodic retraining on the latest data to avoid performance degradation. 
Similarly, there is a question about the robustness of a proxy model trained on a sample of the entire table when applied to specific data slices based on filters. This is an important consideration that needs to be carefully monitored and verified to derive the best sample size and sampling strategy for offline training.

It is also worth noting that for simpler text classification problems, high-quality embedding generation can sometimes be skipped in favor of using a basic word frequency-based classifier, as demonstrated in Table~\ref{tab:simple_binary_classification}.
However, our focus remains on leveraging generally applicable text embedding models due to their robustness and broad applicability across diverse tasks. Furthermore, the proposed proxy model approximation easily extends to non-textual data embeddings \yeounoh{generated with various and multimodal embedding models~\cite{lee2025gemini}.}


\subsection{Imbalanced Label Training}
\label{sec:imbalanced_training_techniques}
The core challenge in training an effective proxy model is the imbalanced label distribution and sourcing sufficient relevant examples from a large table. Label imbalance is a major impediment to model quality and is difficult to completely mitigate with sampling alone. 
There are many techniques to mitigate the problem~\cite{gao2025comprehensive}.
\yeounoh{Namely, these methods include downsampling (reducing the majority class), bootstrapping (resampling of the minority class with replacement), weighted training (assigning higher loss weights to minority class examples, inversely proportional to class frequencies in the training sample.}), and synthetic minority class data generation techniques like, SMOTE (Synthetic Minority Over-sampling Technique~\cite{chawla2002smote}).

Based on some performance and cost analysis, we establish that simple weighted training\footnote{We use \textit{sklearn}~\cite{JMLR:v12:pedregosa11a} Logistic Regression implementation in our prototype pipeline, with \textit{class\_weight = ``balanced''} for weighted training.} provides the most stable, cost-efficient solution, and is the default choice unless there is an insufficient number of minority examples. For all other hyperparameters, we use the default (e.g., L2-Regularization), since tuning does not result in significant differences as discussed in Section~\ref{sec:learning_tuning}.

If there are not enough minority class examples to support proxy model training, then we apply more costly oversampling (SMOTE) technique.
In Section~\ref{sec:imbalanced_training}, we discuss our experimental results, comparing the performance of the various standard imbalanced techniques across various imbalance ratios to validate our heuristics.

\subsection{Automatic Proxy Model Evaluation}
The query engine will automatically evaluate the selected proxy models on a fixed sample from the target dataset. For online training, we label the sample with LLM for training and evaluate the performances of the proxy models on the training sample. If the results do not meet the desired thresholds, as described in Definition~\ref{def:adaptive_proxy}, then we fall back to the LLM baseline -- continue predicting for the rest of the table beyond the initial sample.
For offline training, we can also use pre-curated test datasets, \yeounoh{labeled by LLMs offline}, to evaluate the trained proxy models for known workloads.

\subsection{Adaptive Proxy Model Selection}
\label{sec:adaptive_proxy_model_selection}

\begin{figure*}[t!]
    \centering
    \includegraphics[width=\textwidth,trim={0cm 1.2cm 0cm 3.5cm}, clip]{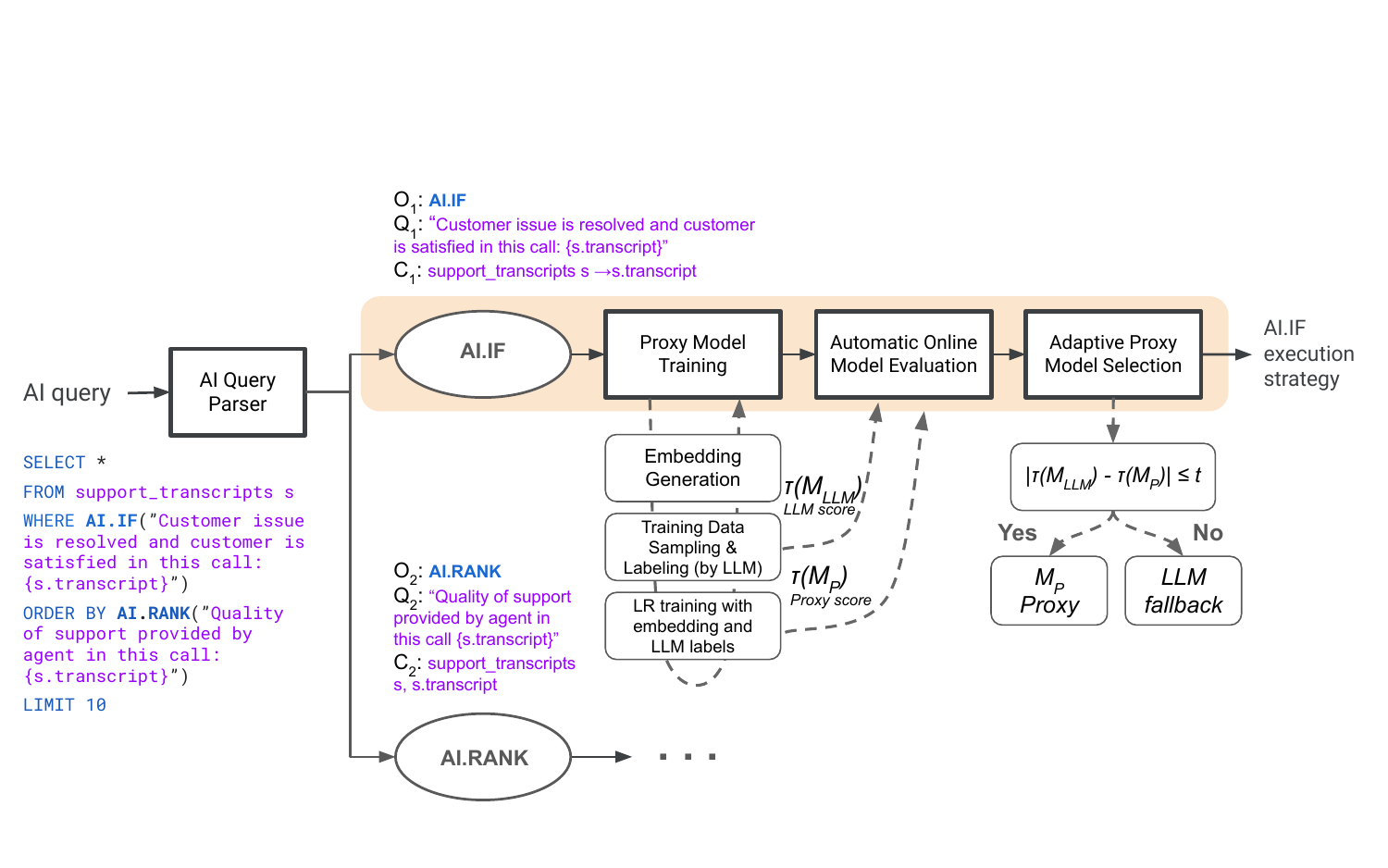}
    \caption{\yeounoh{AI query execution plan construction with proxy model approximation process.} We parse the AI  query and extract semantic operators  ($O_i$) along with the semantic queries/prompts ($Q_i$), unstructured data columns ($C_i$). We apply  proxy model approximation for each operator. Proxy models can be trained online with best known or default configurations, or prepared offline for known query patterns depending on the analytics database architectures (OLAP or HTAP). Adaptive proxy model selection between the proxy and LLM, based on automatic online model evaluation results, allows us to choose a  cost-efficient and accurate execution strategy.}
    \label{fig:ai_query_phys_plan}
\end{figure*}

Table~\ref{tab:simple_binary_classification} illustrates the potential of  simple proxy models performing comparably with the expensive LLM-based baseline operators.
In general, proxy model (also smaller LLMs) substitution for LLM in AI operators requires careful consideration and accuracy guarantees and/or validation as it can compromise quality~\cite{patel2024lotus,shankar2024docetl,palimpzest_cidr2025}. Our approach is to intelligently deploy  offline/online proxy models when applicable and fall back to the LLM-based baselines. 
Namely, we select the proxy if the proxy prediction accuracy is close to the LLM baseline (e.g., $\tau$=0.1, difference is less than absolute 10\% drop from the baseline). We can compute and monitor the proxy accuracy during training using the LLM labels as ground truth.
This adaptive proxy model selection strategy allows us to dynamically balance quality requirements with performance and cost constraints, ensuring the delivery of high-quality semantic insights while operating efficiently for the large-scale analytical workloads.
The problem of adaptive proxy model selection can be formally defined as follows:

\begin{definition}\label{def:adaptive_proxy}
Given an incoming AI query $A$, we extract a set of AI operators $\mathcal{O} = \{O_1, O_2, \ldots, O_m\}$ (e.g., semantic filter).  $A$ also contains the semantic queries $\mathcal{Q} = \{Q_1, Q_2, \dots, Q_m\}$ and the unstructured data column references $\mathcal{C} = \{C_1, C_2, \dots, C_m\}$, used as input to the corresponding AI operators.   
The problem of adaptive model selection is to define a function $\mathcal{S}(O_j, Q_i, C_l) \rightarrow M_p$, where $\mathcal{S}$ is an adaptive selection mechanism that, for any given operator $O_j$, query $Q_i$, and dataset $C_l$, selects an optimal model $M_p \in \mathcal{M}$, where $\mathcal{M} = \{M_{LLM}, M_{P_1}, M_{P_2}, \ldots, M_{P_n}\}$, and $M_{LLM}$ is the full-fidelity, high-cost LLM-based baseline, and $M_{P_1}, \ldots, M_{P_n}$ are various lightweight proxy models. 
$M$ can include both offline and online trainable proxy models. $M_p$ meets the operator specific quality constraint: $|\tau(M_p | O_j, Q_i, C_l) - \tau(M_{LLM} | O_j, Q_i, C_l)| \leq t$ for some relevant samples from $C$ and an expected quality level $\tau$ for the given operator.
If no such $M_p \in \mathcal{M}$ exists, then the operator is served by $M_{LLM}$.
\end{definition}

This definition underpins our approach to adaptive proxy model selection, as depicted in Figure~\ref{fig:ai_query_phys_plan}. We adopt a hierarchical strategy to apply this adaptive selection for each semantic operator. First, given an incoming AI query $A$, our system begins by extracting a set of discrete AI operators (e.g., semantic filter, rank). This decomposition allows us to manage the complexity of multi-operator queries by breaking them down into simpler, actionable components.
For each identified operator, we then implement the selection function S. This function dynamically chooses between $M_P$ (Proxy) and $M_{LLM}$ (LLM) based on the online proxy model evaluation results (e.g., proxy model accuracy over the train sample or any pre-curated test dataset) to deploy the proxy approximation or fall back to LLM-based baseline implementation.

\section{Detailed Analysis of Proxy Approximation}
\label{sec:results}
\begin{table*}[htbp]
    \centering
    \caption{Classification benchmark datasets for AI.IF. For the evaluation of the semantic filter (\textbf{AI.IF}), datasets with multiple categories/labels (\textbf{Relevancy} $\text{>}$ \textbf{Binary}) are used to pose a \textbf{one-vs-rest} semantic query (e.g., matching one specific category against all others).}
    \label{tab:classification_benchmarks}
    \begin{tabularx}{\textwidth}{@{} X p{1.5cm} p{1.5cm} c c c X @{}}
        \toprule
        \textbf{Dataset} & \textbf{Task} & \textbf{Domain} & \textbf{Relevancy} & \textbf{Table (rows)} & \textbf{Imbalance Ratio$^1$} & \textbf{Example Prompt} \\
        \midrule
        California Housing~\cite{Pace1997Sparse} & Text & Housing & Binary & 20K & 6.71 & \footnotesize Location in Latitude \& Longitude is near the ocean. \\
        Amazon Reviews 10k~\cite{Zhang2015Character} & Text & eCommerce & Binary & 10K & 4.69 & \footnotesize Reviews describe at least one issue with the context of the problem. \\
        BBC News~\cite{Greene2006Kernel} & Topic & News & 5 categories & 2.2K & 1.32 & \footnotesize Classify new article into following categories (business / entertainment / politics / sport / tech) \\
        IMDB Movie Reviews~\cite{Maas2011Learning} & Sentiment & Movies & Binary & 99K & 1.10 & \footnotesize Movie review is positive. \\
        Amazon Polarity~\cite{
Zhang2015Character} & Sentiment & eCommerce & Binary & 400K & 1.00 & \footnotesize Review is positive. \\
        Mental Health~\cite{Ji2022MHQA} & Sentiment & Health & Binary & 51.6K & 3.41 & \footnotesize Mental health is anxiety. \\
        Tweet Sentiment~\cite{Kaggle2020TweetSentiment} & Sentiment & SNS & Binary & 31K & 2.21 & \footnotesize Does the following tweet represent positive sentiment? \\
        Emotion \cite{Saravia2018Carer}  & Emotion & SNS & 6 categories & 16K & 9.37 & \footnotesize Can you determine if the emotion expressed in the following statement can be categorized as fear? \\
        Banking77~\cite{Casanueva2020Efficient} & Intent & Finance & 77 categories & 13K & 3.03 & \footnotesize What is the intent of the following banking query? Only return a single intent from the following table, by the \textbf{CODE} (number). \\
        Toxic Conversations~\cite{Wulczyn2017ExMachina} & Text & SNS & Binary & 52K & 11.61 & \footnotesize The conversation is toxic. \\
        \yeounoh{FEVER}~\cite{Thorne18Fever} & \yeounoh{Fact} & \yeounoh{Wikipedia} & \yeounoh{Ternary} & \yeounoh{6.6K} & \yeounoh{1.00} & \yeounoh{\footnotesize Is the claim supported by the text?} \\
        \bottomrule
        \multicolumn{7}{@{} l}{\footnotesize $^1$ Imbalance ratio is defined as the ratio of the number of instances in the majority class to the number of instances in the minority class.}\\
    \end{tabularx}
\end{table*}
\vspace{1em}
\begin{table*}[htbp]
    \centering
    \caption{Information Retrieval (IR) benchmark datasets for AI.RANK. These datasets are designed to test the performance of relevant document retrieval and are typically characterized by heavy class imbalance, where only a handful of relevant documents exist per query within corpora that can contain hundreds of thousands or even millions of documents.}
    \label{tab:ir_benchmarks}
    \begin{tabularx}{\textwidth}{@{} X l l c c l c @{}}
        \toprule
        \textbf{Dataset} & \textbf{Task} & \textbf{Domain} & \textbf{Relevancy} & \textbf{Queries} & \textbf{Documents} & \textbf{Relevant Docs/Query} \\
        \midrule
        TREC-COVID~\cite{Roberts2021SearchingFS} & Bio-medical IR & Bio-medical & 3-level & 50 & 171K & 493.5 \\
        TREC-DL-2022~\cite{Craswell2022OverviewDL} & Ranking/IR & Misc. & 4-level & 500 & 369K & 189.3 \\
        FIQA-2018~\cite{Maia2018FiQA} & Bio-medical IR & Bio-medical & Binary & 648 & 57K & 2.6 \\
        SCIDOCS~\cite{Cohan2020SPECTER} & Argument retrieval & Misc. & Binary & 1000 & 25K & 4.9 \\
        SciFact~\cite{Wadden2020SciFact} & Fact checking & Scientific & Binary & 300 & 5K & 1.1 \\
        HellaSwag$^1$~\cite{zellers2019hellaswag} & Commonsense reasoning & Daily life & Binary & 200 & 800 & 1.0\\
        \bottomrule
        \multicolumn{7}{@{} p{\textwidth} @{}}{\footnotesize $^1$ The task is to find the most commonsensical ending from four plausible-sounding choices for each claim (context). We sample 200 test claims for evaluation, the corpus consists of all possible endings from all the claims. The task is to rank and/or identify the most probable ending among those.}\\
    \end{tabularx}
\end{table*}

The proxy approximation is rigorously evaluated across a wide variety of prompts, questions, and datasets to demonstrate its effectiveness and cost-efficiency.  
Our prototype AI query engine is implemented on two distinct cloud analytics database platforms (\textit{BigQuery} and \textit{AlloyDB}). The \textit{BigQuery} OLAP architecture uses online proxy model training for large-scale online analytics queries. The \textit{AlloyDB} HTAP architecture uses offline trained proxy models for known queries and its ultra fast latency requirements. \yeounoh{AI operators are implemented in the AI query engines as SQL functions and consume table data (transanctionally consistent in HTAP system) as input (just like any other SQL function).}

We evaluate the performance and cost-effectiveness of using proxy models to approximate the semantic filter (AI.IF) and semantic rank (AI.RANK) operators.
For individual micro-benchmarks and experiments in this study, we focus primarily on the AI.IF operator results. This is because, in the context of proxy approximation, the semantic ranking operator also reduces to a classification task based on the input column embeddings -- identical to the case for AI.IF. This means that we can deduce the same approximation success and failure conditions for both operators.
We analyze the proxy's performance across diverse domains, explicitly detailing the conditions under which it succeeds and fails using the benchmark datasets from Table~\ref{tab:classification_benchmarks} and \ref{tab:ir_benchmarks}.

\subsection{Evaluation Setup}\label{sec:eval_setup}
\noindent\textbf{System Implementation and Methodology.}
The choice of Google \textit{BigQuery} and \textit{AlloyDB} cloud databases  
is strategic, designed to cover fundamentally different analytical paradigms. 
\textit{BigQuery} is used for online analytics and leverages massive parallelism for high-throughput, which opens the door to online model training (the extra overhead from sampling, labelling, and training is incurred during query execution).
In contrast, \textit{AlloyDB} targets queries with strict latency requirements, where models are prepared offline for known query patterns. This distinction allows us to isolate the impact of the model preparation methodology. 
To focus on a more comprehensive scenario, involving the proxy model training process, we primarily report results from \textit{BigQuery} (OLAP), providing a latency breakdown and model performance metrics that explicitly account for the overhead of the necessary online training. \textit{AlloyDB} (HTAP) removes this training process offline for known queries and workload patterns, effectively removing the overhead from the execution critical path. We also report separate normalized latency gains from \textit{BigQuery} and \textit{AlloyDB} highlighting some key system architecture differences in Section~\ref{sec:filtering_results}. 

In our production engine implementation, while the execution of the sampling, labeling, and prediction steps is distributed, we opted to execute the training step for the logistic regression serially to minimize communication and scheduling overheads.
For proxy model training, we primarily utilize the Gecko embedding model (\textit{text-embedding-005})~\cite{Lee2024Gecko} \yeounoh{to compute the embeddings of the unstructured text data columns and semantic queries}, except when we study the impact of various embedding models and dimensionality in Section~\ref{sec:embedding_quality}. 
\yeounoh{All proxy model training and inference were executed solely on standard vCPUs (e.g., 8 vCPU, 16 GB RAM instances), and we did not use HW accelerators (e.g., GPUs). This design choice validates our core claim: the proxy models  are sufficiently lightweight to deliver high performance on commodity hardware without requiring expensive HW acceleration.}

\noindent\textbf{Baselines:} Our primary baseline for evaluating the performance and cost-effectiveness of the proxy approximation is a pure Large Language Model (LLM) execution of the same semantic operators, specifically utilizing the one of the most capable LLM family (e.g., Gemini~\cite{GoogleDeepMind2025Gemini})\footnote{We use Gemini LLM (2.5-Flash) for the evaluation; however, our approach is fundamentally LLM-agnostic as we examine the feasibility of substituting LLM baseddiscriminative semantic task (filter and rank) with non-LLM based ML classifier.}.  This pure LLM based baseline is deemed to be the most accurate and also the most expensive baseline that other contemporary LLM-centric semantic query systems try to optimize~\cite{jo2024thalamusdb,patel2024lotus,russo2025abacus,shankar2024docetl}. 

Recent AI query optimization techniques employs predictive row skipping, filter re-ordering and/or model cascading to reduce the expensive LLM calls. As our focus is to evaluate proxy model and the possibility of processing the entire table with millions of rows, we do not compare directly with other optimization techniques.
For a reference, \cite{patel2024lotus} sampled queries and corpus documents (search space) to below 1K rows  ``to control for our API spending budget'' in their evaluations.

\noindent\textbf{Datasets.} For semantic filtering (AI.IF), we implement a binary classifier using logistic regression proxy. These models are evaluated using the Classification benchmark datasets ($\text{Table~\ref{tab:classification_benchmarks}}$).
For semantic ranking (AI.RANK), we consider both the logistic regression proxy and the state-of-the-art cross-attention re-ranker to provide high-quality ranking results with low latency and cost. These models are evaluated using the Information Retrieval (IR) benchmark datasets ($\text{Table~\ref{tab:ir_benchmarks}}$).

\subsection{Fast Semantic Filtering}
\label{sec:filtering_results}

\begin{table*}[!ht]
\centering
\small 
\setlength{\tabcolsep}{4pt} 
\caption{Macro F1 and relative accuracy of the proxy vs. LLM baseline for semantic filtering. For multi-label datasets, results are averaged over one-vs-rest binary classification tasks.}
\label{tab:ai_if_results}
\begin{tabularx}{\textwidth}{
    >{\raggedright\arraybackslash}p{2.6cm} 
    S[table-format=2] 
    >{\raggedright\arraybackslash}X 
    S[table-format=1.3] 
    S[table-format=1.3] 
    S[table-format=1.3] 
}
\toprule
\textbf{Dataset} & {\makecell[c]{\textbf{Labels}}} & \textbf{Per-label Prompt (Binary)} & {\makecell[c]{\textbf{Macro F1} \\ \textbf{Proxy}}} & {\makecell[c]{\textbf{Macro F1} \\ \textbf{LLM}}} & {\makecell[c]{\textbf{Relative} \\ \textbf{Acc.}$^1$}} \\
\midrule
California Housing & 1 & \footnotesize{Location in Latitude \& Longitude belongs to Southern California.} & \cellcolor{myLightGreen} 0.953 & \cellcolor{myLightGreen}0.953 & \cellcolor{myLightGreen}\textbf{1.000} \\
California Housing & 1 & \footnotesize{Location in Latitude \& Longitude is near the ocean.} & \cellcolor{myLightRed}0.344 & \cellcolor{myLightRed}0.354 & \cellcolor{myLightGreen}\textbf{0.971} \\
Amazon Reviews 10k & 2 & \footnotesize{Review is \{sentiment label\}} & 0.860 & 0.739 &\cellcolor{myLightGreen}\textbf{ 1.163} \\
BBC News & 5 & \footnotesize{News Content describes news in \{news topic label\} sector} & \cellcolor{myLightGreen}0.830 & \cellcolor{myLightGreen}0.823 & \cellcolor{myLightGreen}\textbf{1.008} \\
IMDB Movie Reviews & 2 & \footnotesize{Movie review is \{sentiment label\}} &\cellcolor{myLightGreen} 0.940 & \cellcolor{myLightGreen}0.950 & \cellcolor{myLightGreen}\textbf{0.989} \\
Amazon Polarity & 2 & \footnotesize{Review is \{sentiment label\}} & \cellcolor{myLightGreen}0.965 & \cellcolor{myLightGreen}0.959 &  \cellcolor{myLightGreen}\textbf{1.005}\\
Mental Health & 1 & \footnotesize{Mental health described by the patient relates to anxiety or stress:} & \cellcolor{myLightRed}0.214 & \cellcolor{myLightRed}0.230 & \cellcolor{myLightGreen}\textbf{0.931} \\
Mental Health & 1 & \footnotesize{Mental health is normal} & \cellcolor{myLightRed}0.357 & \cellcolor{myLightRed}0.349 & \cellcolor{myLightGreen}\textbf{1.022} \\
Tweet Sentiment & 2 & \footnotesize{Does the following tweet represent \{sentiment label\} sentiment?} & \cellcolor{myLightGreen}0.880 & \cellcolor{myLightGreen}0.890 & \cellcolor{myLightGreen}\textbf{0.989} \\
Emotion Classification & 6 & \footnotesize{Can emotion described can classify as \{emotion label\} emotion?} & \cellcolor{myLightRed}0.428 & \cellcolor{myLightRed}0.475 & \cellcolor{myLightGreen}\textbf{0.901} \\
Toxic Conversations & 2 & \footnotesize{The conversation is toxic.} & 0.653 & 0.648 & \cellcolor{myLightGreen}\textbf{1.006} \\
Banking77 & 77 & \footnotesize{Is intent \{intent label\}? Think step-by-step: \{CoT instructions\}.} & 0.700 & 0.707 & \textbf{0.990} \\
\yeounoh{FEVER} & \yeounoh{2$^2$} & \yeounoh{\footnotesize{Is the claim supported by the text?}} & \yeounoh{0.782} & \yeounoh{0.853} & \yeounoh{ \textbf{0.917}} \\
\bottomrule
\multicolumn{6}{@{} p{\textwidth}}{\footnotesize $^1$ Relative accuracy is the ratio between Proxy and LLM Macro F1 scores.}\\
\multicolumn{6}{@{} p{\textwidth}}{\footnotesize $^2$ \yeounoh{We follow \cite{patel2024lotus} and \cite{chern2023factool} by merging ``Refuted'' and ``NotEnoughInfo'' into a single ``Not Supported'' category.}}
\end{tabularx}
\end{table*}

In terms of quality, we obtain high F1 scores\footnote{F1-score is defined as harmonic mean of precision and recall. F1-score is a better metric than mere accuracy since many of the datasets are highly imbalanced.} across the various benchmark datasets as presented in Table~\ref{tab:ai_if_results}. 
To test AI.IF semantic filtering query over benchmark datasets with multiple label classes, we pose a binary classification query per label and take the average of per label F1-score as macro F1-score.
It is important to note that the proxy approximation yields very high (near perfect) relative accuracy (the ratio between macro F1 Proxy and macro F1 LLM), indicating that the proxy model predictions can closely match the LLM; in some cases (e.g., Amazon Reviews 10k, BBC News, Mental Health, Toxic Conversation) the proxy approximation performs slightly better (relative accuracy $\geq$ 1.0), since the simple and low-complexity LR decision function can act as a regularizer on top of embeddings.
Given that the proxy model learns and approximates the LLM predictions, the approximation also fails when the LLM predictions are inaccurate -- for very difficult semantic filtering queries. 

\yeounoh{On the FEVER dataset, our proxy approach achieves a 0.917 relative accuracy compared to the LLM baseline while processing the entire 6.6k claim set -- the inference takes just a few seconds, and the training pipeline, including 1,000-sample sampling, labeling, and model fitting takes less than a minute with a scale-out system. 
In contrast, plan-optimization frameworks like LOTUS~\cite{patel2024lotus} report significantly higher latencies (e.g., $\sim$190s per 1,000 claims). 
While hardware variations preclude a direct comparison, these results highlight the distinct cost-efficiency of proxy substitution. 
Unlike LOTUS and DocETL~\cite{shankar2024docetl}, which focus on optimizing execution plans and reducing redundant LLM calls, our approach aims to eliminate LLM inference costs much more aggressively for high-frequency operations. These strategies are orthogonal; integrating proxy-based approximation into plan-optimized architectures could yield synergistic reductions in both cost and latency.}

\begin{table}[!ht]
\begin{minipage}{\columnwidth}
\centering
\caption{Cost and latency improvement of online proxy model approximation VS. the LLM baseline, with (On-the-Fly) and without (Pre-Computed) embeddings, tested using Amazon Polarity.}
\label{tab:embedding_comparison}
\resizebox{0.95\columnwidth}{!}{
\begin{tabular}{l|c|c|c|c}
\toprule
 & \multicolumn{2}{c|}{\textbf{On-The-Fly}} & \multicolumn{2}{c}{\textbf{Pre-Computed}} \\
\cmidrule(lr){2-3} \cmidrule(lr){4-5}
\textbf{Table Size} & \textbf{Cost $\uparrow$} & \textbf{Latency $\uparrow$} & \textbf{Cost $\uparrow$} & \textbf{Latency $\uparrow$} \\
\midrule
10,000 & 2x & 1.3x & 9x & 1.6x \\
100,000 & 2.5x & 3x & 81x & 6x \\
1,000,000 & 2.5x & 5x & 422x & 30x \\
10,000,000 & 2.5x & 5x & \textbf{728x} & \textbf{320x} \\
\bottomrule
\end{tabular}
}
\end{minipage}
\end{table}

In terms of scalability and costs, we obtain orders of magnitude speed-ups and cost savings over our pure LLM baseline with pre-computed embeddings~(Table~\ref{tab:embedding_comparison}). With proxy-models, AI Queries become practical in OLAP workloads that are typical of databases like \textit{BigQuery}. 
The latency gain scales superlinearly as larger tables (more rows) gives more opportunity for \textit{BigQuery} to parallelize execution. Even with proxy models, the costs are primarily driven by LLM invocation in the labeling step. Since the number of samples to label is independent of the input table size, the reduction in cost also scales superlinearly with the input rows.
Our results also show that a pure online version (On-The-Fly Embedding) of proxy models is worthwhile for OLAP workloads. When we generate embeddings on-the-fly for each query (with no reuse or amortization of embedding generation), we achieve 5x speed-up and 2.5x reduction in cost. The 2.5x reduction in cost remain constant for all data sizes because it reflects the fact that embedding generation is roughly a third of the cost of LLM inference. (This is not a perfect comparison as embedding generation is charged per character, while LLM inference is per token).

\begin{table}[!ht]
\begin{minipage}{\columnwidth}
\centering
\caption{Cost and latency improvement of offline proxy model approximation VS. the LLM baseline, tested using Amazon Polarity. Offline training includes the same processing (vCPU), labeling and embedding costs as in Table~\ref{tab:embedding_comparison} per the same train sample size.}
\label{tab:embedding_comparison2}
\resizebox{0.6\columnwidth}{!}{
\begin{tabular}{l|c|c}
\toprule
\textbf{Table Size} & \textbf{Cost $\uparrow$} & \textbf{Latency $\uparrow$} \\
\midrule
10,000 & 9x & \multirow{4}{*}{\textbf{111x}} \\
100,000 & 81x & \\
1,000,000 & 422x & \\
10,000,000 & 728x & \\
\bottomrule
\end{tabular}
}
\end{minipage}
\end{table}

Our \textit{AlloyDB} HTAP architecture with offline proxy model training is designed for latency-sensitive database applications. It uses pre-trained offline proxy models prepared for known workload patterns and with a fixed number of workers in a single database VM instance to ensure high latency and throughput gain even for smaller table sizes. The latency gain against the LLM baseline is constant across different table sizes, as shown in Table~\ref{tab:embedding_comparison2}. The fixed gain is driven by the stark difference in the LLM and proxy model inference times and the fixed number of parallel workers, which is also configurable; the processing time for both the LLM baseline and the offline proxy are linear in the size of table.

\begin{figure}[!ht]
    \centering
    \includegraphics[width=\columnwidth,height=0.55\columnwidth]{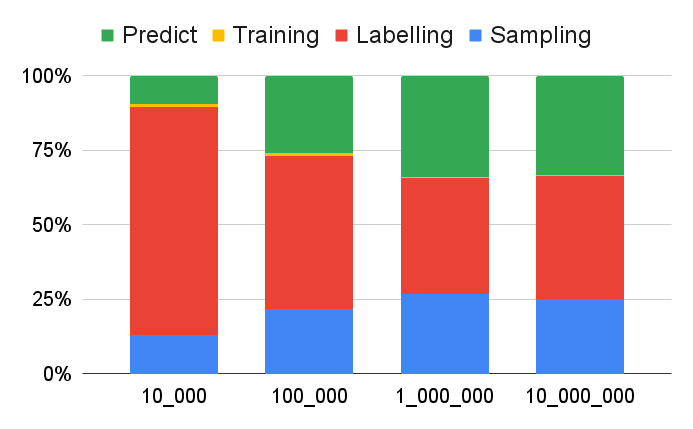}
    \caption{
    Relative wall-clock time of each step of the proxy model optimization with pre-computed embeddings.
    Only the sampling and prediction steps scale with the input relation size. The fraction of time spent doing prediction does not increase between 1m and 10m rows because \textit{BigQuery} is able to use more parallelism as the amount of data increases.
    }
    \label{fig:filter_cost}
\end{figure}

Figure~\ref{fig:filter_cost} shows the relative time spent in each step. The cost of model training is insignificant even at small data size. The small relative cost of sampling and training points to the possibility of applying the proxy model optimization eagerly and falling back dynamically on LLM evaluation after evaluating the proxy model accuracy. 
That is, the sunk cost of such a fallback consists only of the sampling and training steps because the labels used for training can be reused.
The overhead of the futile proxy model training is mostly sampling which is kept at ~25\%; this overhead can be further reduced with more parallel workers.

\subsection{Fast Semantic Ranking}
\label{sec:fast_rubric_ranking}

For general ranking tasks, the state-of-the-art cross-attention re-ranker on Vertex AI ~\cite{bruderer2025boost} 
offers high-quality semantic ranking with very low latency based on embedding vectors. However, as ranking queries become more complex (e.g., rubric-based), nuanced LLM-based ranking algorithms may outperform the re-ranker model~\cite{sharifymoghaddam2025rankllm,sun2023chatgpt,patel2024lotus}.
To reduce dependence on expensive LLMs for the more challenging rubric-based ranking tasks without sacrificing ranking quality, we explore applying the proxy approach based on embeddings and simple classification algorithms. 

\begin{figure}[!th]
    \centering
    \includegraphics[width=\columnwidth]{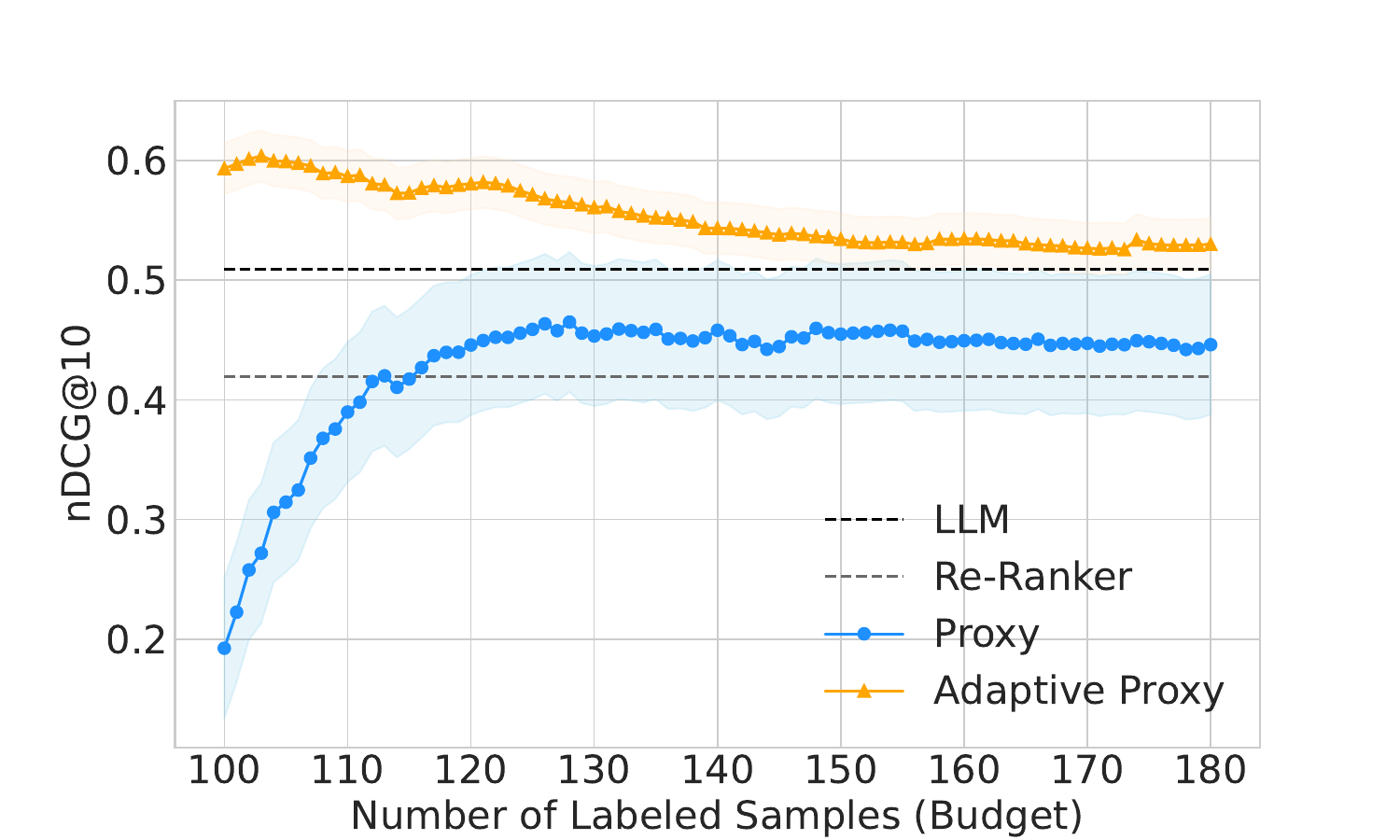}
    \caption{Proxy model performance (nDCG@10 on online training sample) for rubric-based ranking on TREC-DL-2022. The online Proxy performs better after labeling >120 samples; Adaptive Proxy relies on LLM until the Proxy becomes as good as the baseline.}
    \label{fig:fast_rubric_ranking}
\end{figure}

To enable aggressive cost savings for semantic ranking, we reformulate the task as a classification problem. Traditional Information Retrieval (IR) simplifies it to a binary classification of relevant vs. irrelevant. For rubric-based ranking, the scoring rubrics/guidelines (e.g., 4-level rubric in TREC-DL-2022) provide granular classification labels.
The adaptive proxy model selection mechanism then chooses an LLM-based ranking only when the quality constraint is not met: $|\tau(M_p | O_j, Q_i, C_l) - \tau(M_{LLM} | O_j, Q_i, C_l)| \leq t$ (Definition~\ref{def:adaptive_proxy}). Figure~\ref{fig:fast_rubric_ranking} illustrates that we can train an effective proxy model using embeddings and multi-label classifier for rubric-based ranking (TREC-DL-2022~\cite{Craswell2022OverviewDL}), and then Adaptive Proxy can choose between the LLM baseline and the online proxy model. 
\yeounoh{The ranking performance is measured by normalized Discounted Cumulative Gain at rank 10 (nDCG@10). nDCG@10 measures the quality of the top 10 results by considering both the relevance of each item and its position in the ranked list.}

\begin{table}[th]
\centering
\caption{Semantic ranking performance (nDCG@10) on the IR benchmark datasets from Table~\ref{tab:ir_benchmarks}. We report the average nDCG@10 over all queries ($\pm$ 1-std). TREC-DL-2022 ranks items by 4-level  scoring rubric, for which LLM and its Proxy perform better than Re-Ranker.}
\label{tab:ir_benchmark_results}
\resizebox{\columnwidth}{!}{
\begin{tabular}{l|c|c|c}
\toprule
& \textbf{Re-Ranker} & \textbf{LLM} & \textbf{Proxy} \\
\midrule
FIQA-2018
& \textbf{0.432} ($\pm$ 0.363) 
& 0.0703 ($\pm$ 0.107) 
& 0.0304 ($\pm$ 0.027) \\
SciFact
&  \textbf{0.746} ($\pm$ 0.368) 
&  0.508 ($\pm$ 0.416) 
&  0.0098 ($\pm$ 0.009) \\
SCIDOCS 
& \textbf{0.251} ($\pm$ 0.267) 
&  0.107 ($\pm$ 0.151) 
& 0.0185 ($\pm$ 0.0469) \\
TREC-DL-2022 
&  0.419 ($\pm$  0.245) 
&  \textbf{0.537} ($\pm$  0.164) 
&   0.446 ($\pm$  0.241) \\
TREC-COVID
&   \textbf{0.738} ($\pm$ 0.232) 
& 0.551 ($\pm$ 0.167) 
& 0.535 ($\pm$ 0.215) \\
\yeounoh{HellaSwag}
&   \textbf{0.617} ($\pm$ 0.252) 
& 0.247 ($\pm$ 0.306) 
& 0.134 ($\pm$ 0.251) 
\\
\bottomrule
\end{tabular}
}
\end{table}

\begin{table}[ht]
\centering
\caption{Semantic ranking cost and latency comparison (1x for Proxy) for ranking 500 candidate documents. Proxy requires labeling 200 documents with LLM for training.}
\label{tab:ir_benchmark_cost_savings}
\resizebox{0.65\columnwidth}{!}{
\begin{tabular}{l | c | c |c}
\toprule
  & \textbf{Re-Ranker} & \textbf{LLM} & \textbf{Proxy}\\
\midrule
Cost & $0.025\times$ & $2.5\times$ & $1\times$ \\
Latency & $0.11\times$ & $1.1\times$ & $1\times$ \\
\bottomrule
\end{tabular}}
\end{table}

Table~\ref{tab:ir_benchmark_results} summarizes the retrieval and re-ranking performances using the state-of-the-art cross-attention re-ranker~\cite{bruderer2025boost} (Re-Ranker), the LLM-based semantic ranking (LLM) and its simple logistic regression and rich text embedding-based proxy approximation (Proxy) across various IR benchmark datasets from Table~\ref{tab:ir_benchmarks}. To control the cost of ranking tens and hundreds of documents per query, we filter 500 candidate documents per query based on the query and document embedding vector similarity before re-ranking to retrieve the top-10 documents ordered by the relevancy; 
the proxy model is trained on 200 random samples (labeled by LLM) from the 500 candidates, successfully eliminating 300 LLM calls per query.
The similarity based pre-filtering (Top-K) is a standard practice for semantic ranking to control the cost in practice, as well as in the literature~\cite{patel2024lotus,nouriinanloo2024re}. Such sampling strategy is still more expensive than RANDOM (the default sampling for semantic filtering) but much cheaper than Stratified, as shown in Table~\ref{tab:sampling_overheads}.

Overall, Re-Ranker achieves the best performance on five out of six datasets. 
For a candidate set of 500, it is also the most efficient option, operating at just 2.5\% of the proxy's normalized cost\footnote{Re-Ranker can batch 100 documents per inference (5 inference calls to rank 500 documents) without quality degradation, whereas we need 200 LLM inference calls to label the training data for Proxy.} and 1\% of the LLM baseline, as shown in Table~\ref{tab:ir_benchmark_cost_savings}.
Re-Ranker underperformed on TREC-DL-2022, which requires nuanced 4-level scroing rubrics. In such reasoning-heavy tasks, the LLM provides superior quality, and Proxy acts as a high-fidelity approximation at roughly 40\% of the baseline LLM cost.

\yeounoh{The HellaSwag results highlight the limits and potential of proxy-based ranking. Unlike simpler relevance-based ranking, HellaSwag requires nuanced commonsense reasoning to select the most sensical ending out of four possible options for a given context. Despite this difficulty, the proxy provides a  latency reduction, corresponding to 300 LLM inference calls, while maintaining a nDCG@10 within 54\% (relative accuracy) of the LLM. Although HellaSwag benchmark has a moderate class label imbalanced ratio, 1 positive (true ending) to 3 negative (false endings) per claim, we always pre-filter 200 training samples for training, leading to heavy imbalanced ratio with just a single positive instance.} When there is insufficient minority class examples, the proxy model cannot learn the correct LLM decision functions.
Proxy fails to learn the LLM decision functions for the similar reasons on SciFact, SCIDOCS, FIQA-2018, where the number of relevant documents per query ranges from 1.1 to 4.9, making effective sampling and training near impossible.
In these instances, the imbalance ratio increases linearly with sample size (Figure~\ref{fig:impact_of_sampling}(c)), necessitating an automatic fallback to the LLM baseline.

While we focus on evaluating the proxy model approach as an substitution for LLM-based AI.RANK, the results strongly indicate that the existing state-of-the-art cross-attention re-ranker (Re-Ranker) is very effective in many semantic ranking settings~\cite{bruderer2025boost,iyer2021reconsider}.
Given these results, we identify the integration of specialized re-rankers into our adaptive selection framework as a key future direction (Section~\ref{sec:adaptive_ranker_selection}).

\subsection{Impact of Sampling Strategy}
\label{sec:impact_of_sampling}
The practice of sampling is such a crucial step to the efficacy and the accuracy of the proxy approximation. Ideally, the training sample for proxy models should be sufficient and well balanced in class labels to ensure stability and accuracy in predictions. As prescribed in ~\cite{dai2024uqe}, active learning based sampling can greatly improve the sample efficiency for proxy model training (i.e., achieve the same model quality with fewer samples). Furthermore, class label balanced sampling is critical for the accuracy of proxy approximation because a proxy model trained with heavily imbalanced data yields poor performance (Section~\ref{sec:imbalanced_training}).

\begin{figure*}[!th]
    \centering
    \subfloat[$\rho$=11.61]{\includegraphics[width=0.24\textwidth, trim=0.2cm 0.2cm 0.2cm 0.2cm,clip]{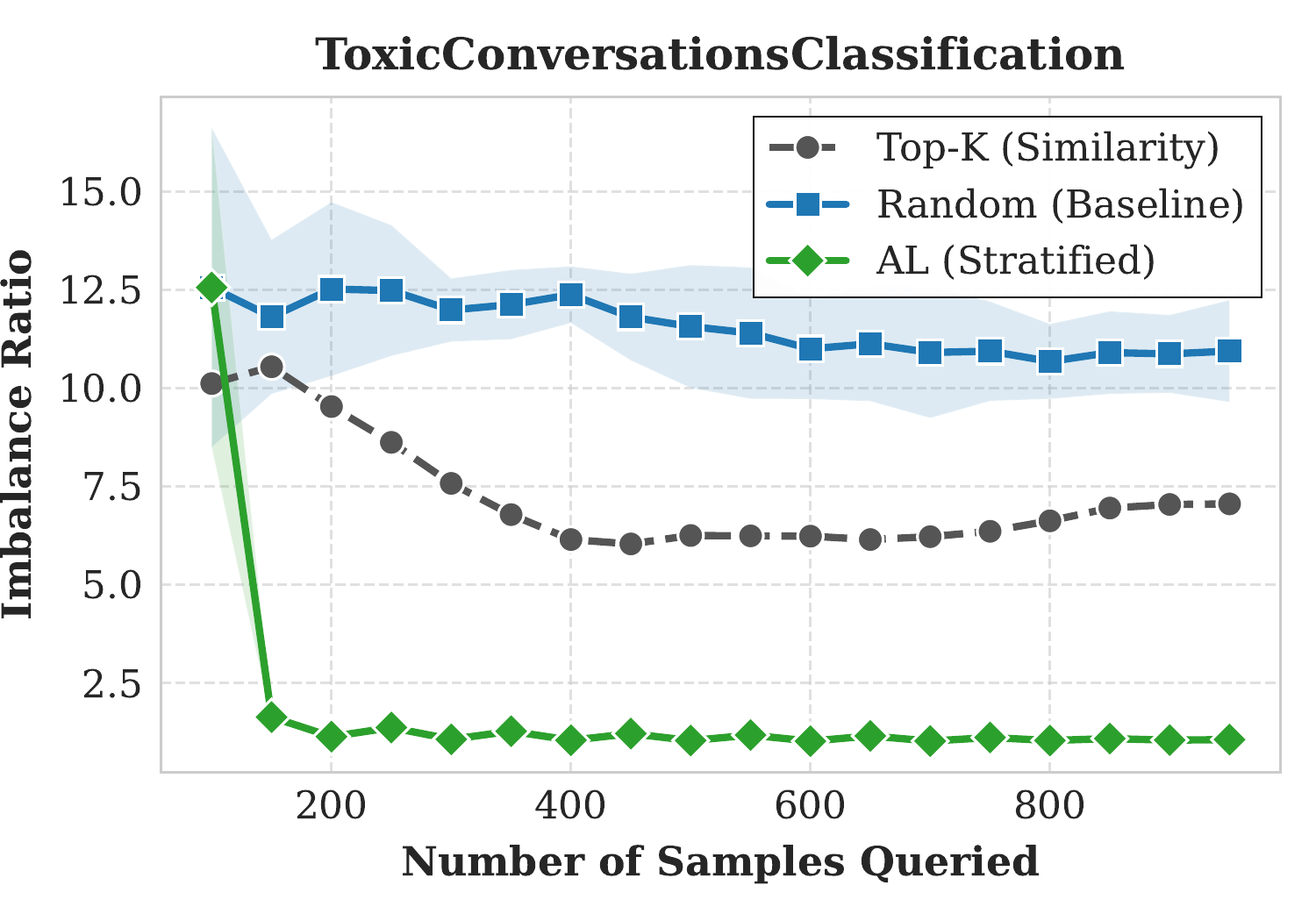}}
    \hfill
    \subfloat[$\rho$=2.21]{\includegraphics[width=0.24\textwidth, trim=0.2cm 0.2cm 0.2cm 0.2cm,clip]{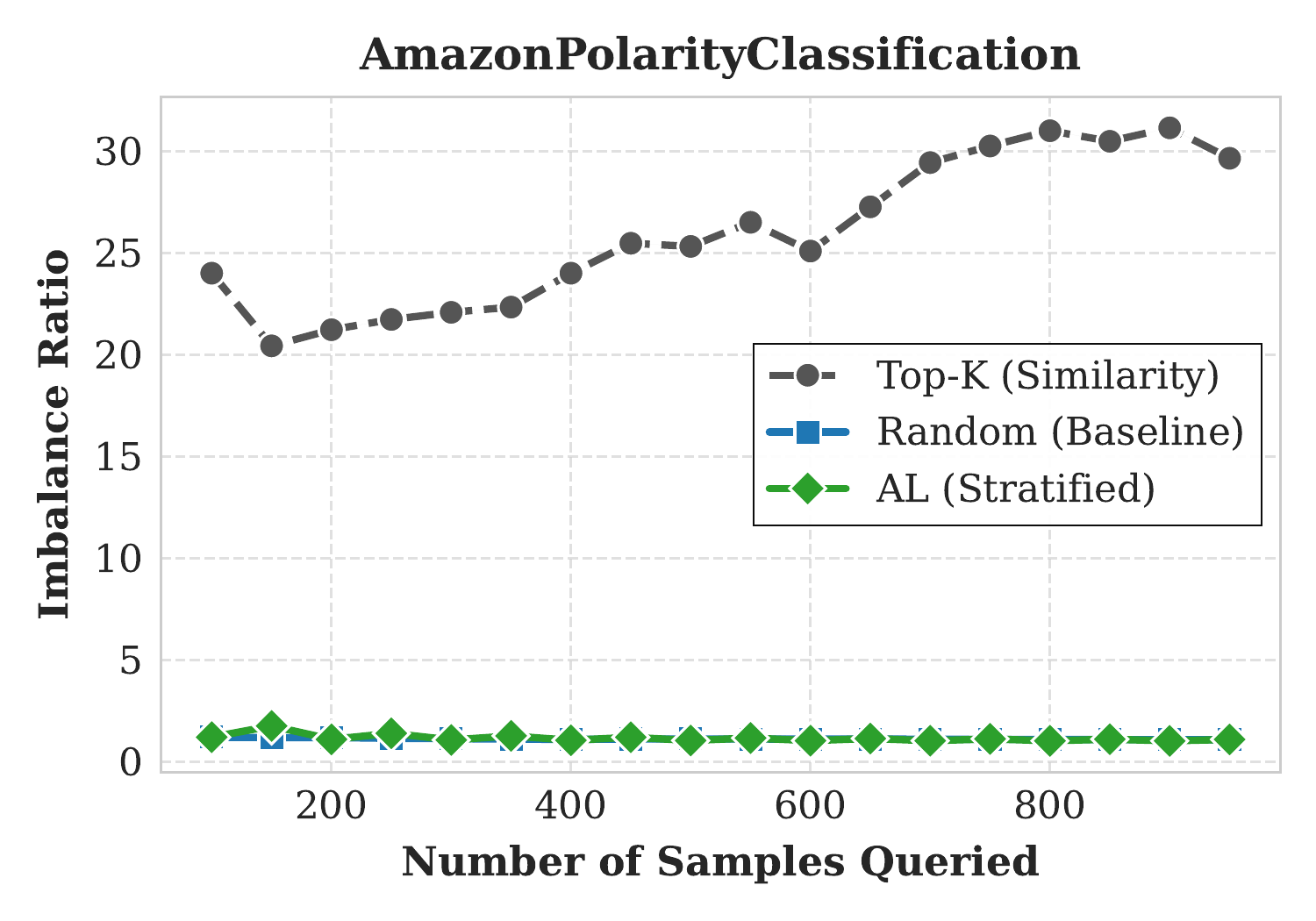}}
    \hfill
    \subfloat[$\gamma$=1.1/5K]{\includegraphics[width=0.24\textwidth, trim=0.2cm 0.2cm 0.2cm 0.2cm,clip]{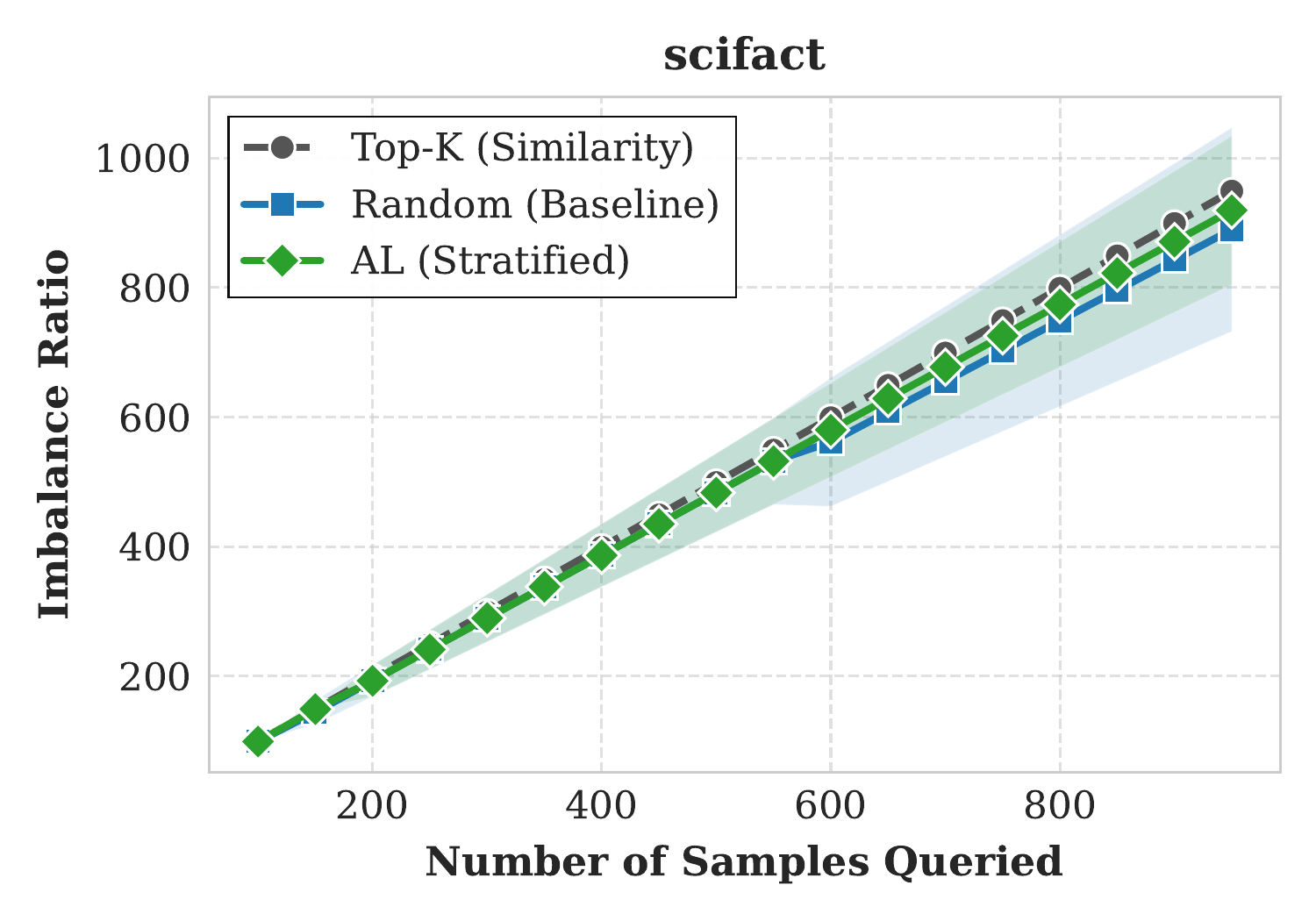}}
    \hfill
    \subfloat[$\gamma$=493.5/171K]{\includegraphics[width=0.24\textwidth, trim=0.2cm 0.2cm 0.2cm 0.2cm,clip]{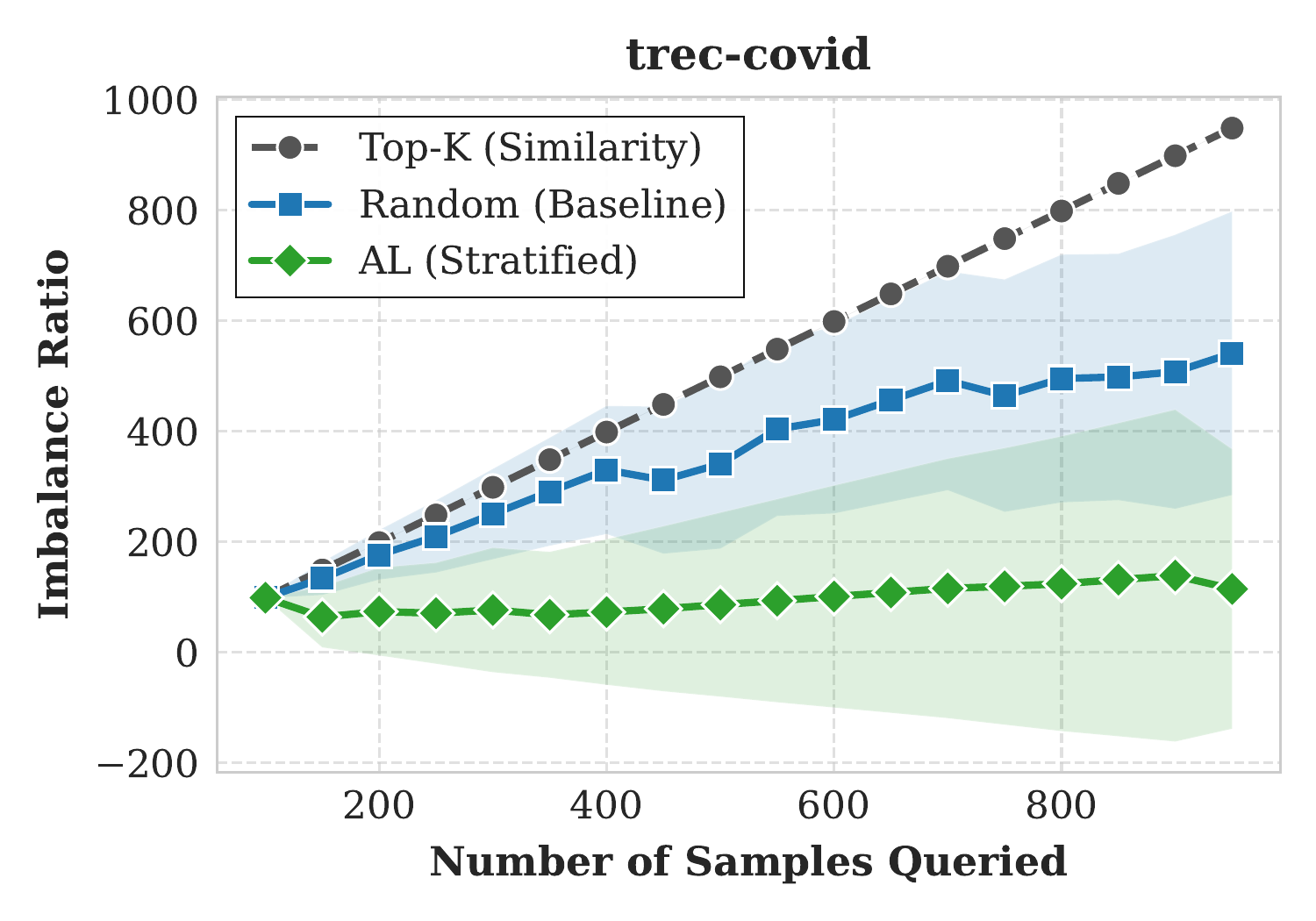}}
    \caption{Impact of sampling strategies on training data imbalance ratios, measured across various datasets of varying degrees of dataset characteristics: imbalance ratio ($\rho$), for classification benchmark datasets (a)(b); Relevant docuemtns per query ($\gamma$) / the number of corpus documents, for IR benchmark datasets (c)(d).}
    \label{fig:impact_of_sampling}
\end{figure*}

Figure~\ref{fig:impact_of_sampling} illustrates how different sampling strategies perform in terms of sample imbalance ratios, across different benchmark datasets. We use two classification benchmark datasets of low and high population imbalance ratios ($\rho$) and two IR benchmark datasets of low and high relevant docs/query ratios ($\gamma$) -- high imbalance ratio and low relevant docs/query mean that the datasets are much harder to sample from to obtain a well balanced training sample.
We consider three sampling strategies: top-K similarity based sample retrieval (Top-K), random sampling (Random), active learning (AL) for stratified sampling. 
AL takes the proxy model prediction confidence to identify most-likely majority or minority class examples and always samples the minority class examples.
On the one hand, when the population label distribution is highly imbalanced (a), both Top-K and Random fail to establish a good balanced training dataset. On the other hand, (b) low population imbalance ratio allows the cheap Random to perform as good as  more expensive stratified sampling (AL).
IR datasets are always heavily imbalanced due to much larger corpus document size.
For this reason, retrieval is relatively harder to apply our proxy model approximation process, as discussed in Section~\ref{sec:fast_rubric_ranking}. Here we look at two contrasting cases with $\gamma$=1.1 and $\gamma$=493.5.
When there is just one relevant document out of 5K documents (c), it is near impossible to establish a meaningful and balanced training sample, even with oversampling. Consequentially, the imbalance ratio almost linearly increases with the number of samples.
When there are a sufficient number of relevant documents per query (d), we can build meaning training sample with the stratified sampling strategy (AL). Note that Top-K results in higher imbalance ratios as it blindly favors one class over the other by query and document relevancy.

\begin{table}[ht]
\centering
\caption{Latency overhead multipliers (1$\times$ for Random) for sampling 1000 sample instances from the Toxic Conversations dataset with 52K rows.}
\label{tab:sampling_overheads}
\resizebox{0.6\columnwidth}{!}{
\begin{tabular}{c | c | c }
\toprule
 \textbf{Random} & \textbf{Top-K} & \textbf{AL (Stratified)}  \\
\midrule
$1\times$ & $43761\times$ & $51639\times$ \\
\bottomrule
\end{tabular}}
\end{table}

Table~\ref{tab:sampling_overheads} illustrates the stark cost difference between Random and AL (Stratified). 
Note that the sampling step can be parallelizable over multiple workers and shards. \yeounoh{Here, we measure the serialized latency in order to focus on the sampling overhead at the algorithm-level.}
While more advanced stratified sampling (AL) can yield more robust and balanced training sample, simple random sampling can also perform well if the dataset (table) is not heavily imbalanced.

\subsection{Imbalanced Data Label Challenge}
\label{sec:imbalanced_training}

\begin{figure*}[!th] 
    \centering
    \subfloat[Amazon Polarity]{\includegraphics[width=0.33\textwidth,]{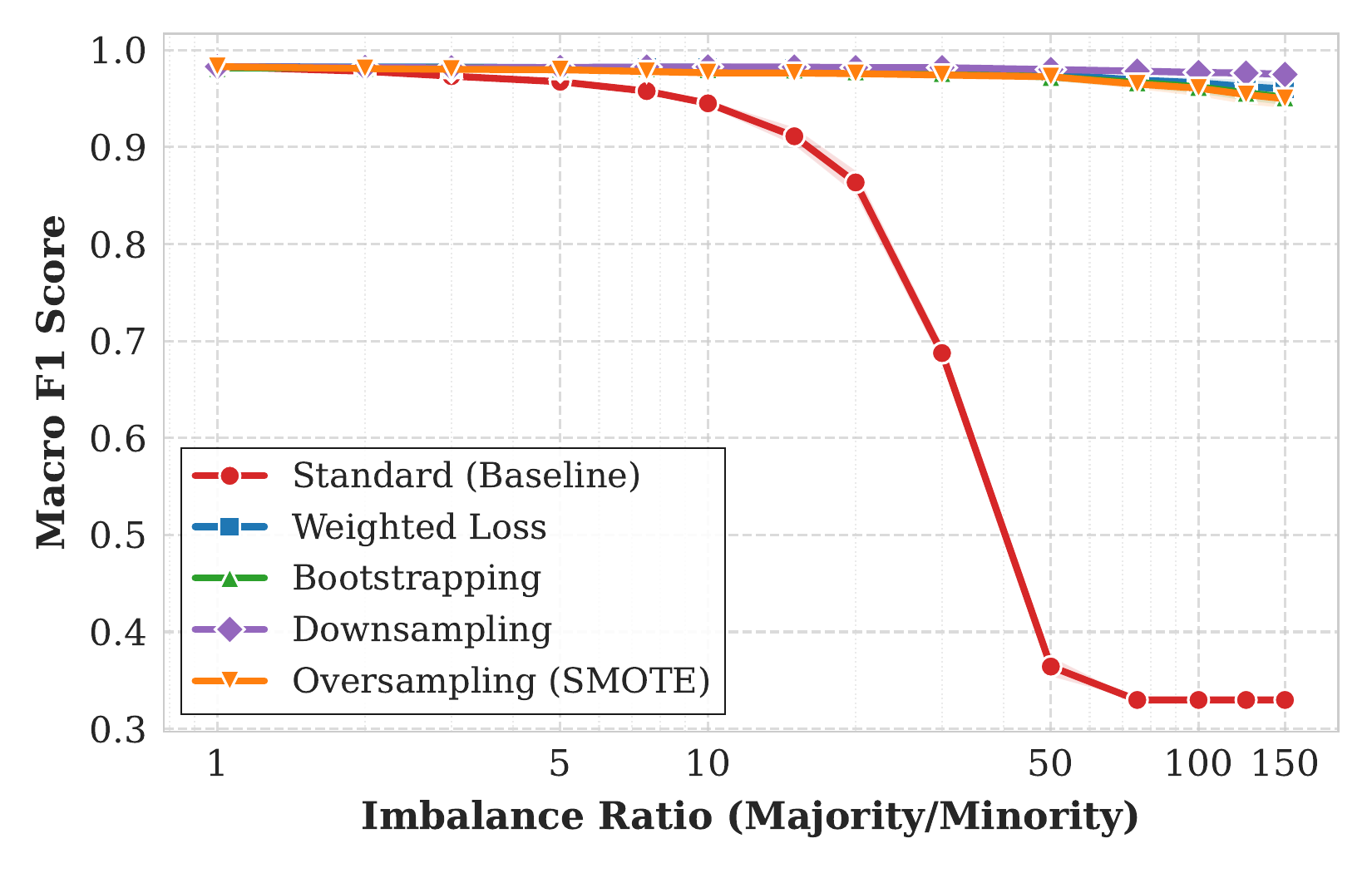}}
    \hfill
    \subfloat[Tweet Sentiment]{\includegraphics[width=0.33\textwidth]{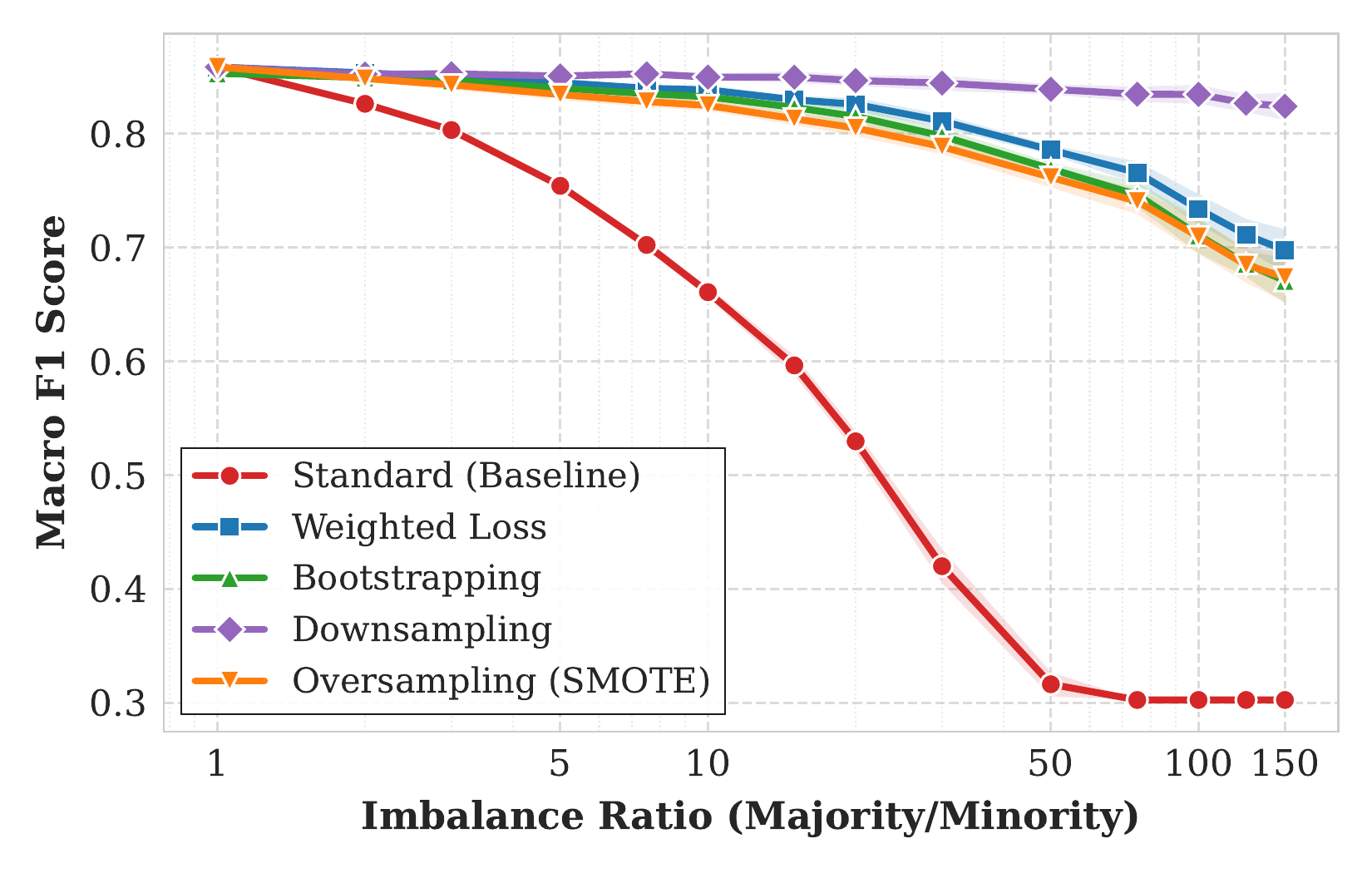}}
    \hfill
    \subfloat[Toxic Conversation]{\includegraphics[width=0.33\textwidth]{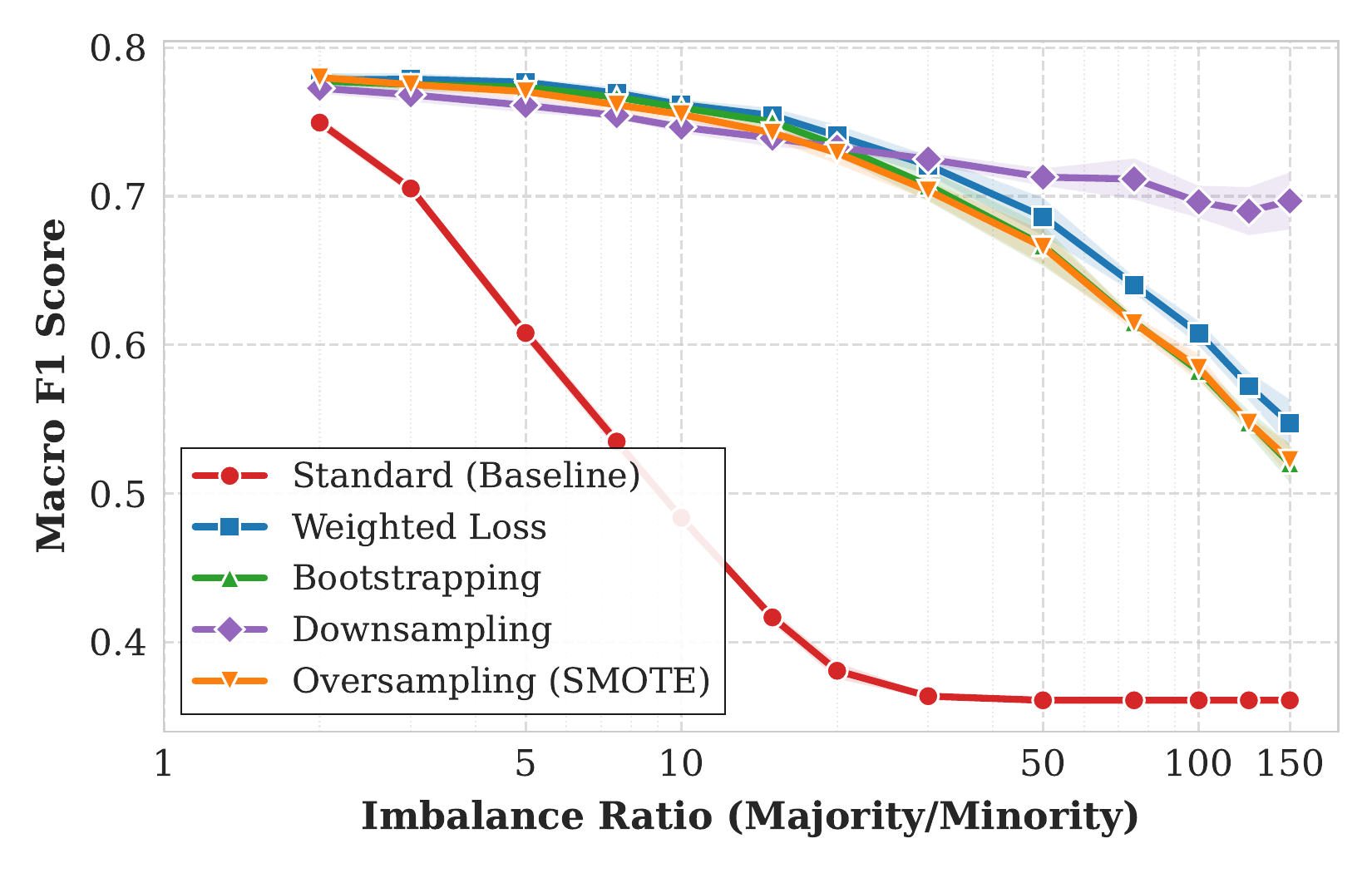}}
    \caption{Impact of different imbalanced label training techniques, as described in Section~\ref{sec:imbalanced_training_techniques}.}
    \label{fig:imbalanced_training}
\end{figure*}

The performance of the original imbalanced trained (Standard) is not robust against the data label imbalance, whereas various imbalanced training techniques can significantly increase model robustness and improve generalization across different imbalance ratios, as illustrated in Figure~\ref{fig:imbalanced_training}.
With heavily skewed label distributions (imbalance ratio $\geq$ 50.0), simple downsampling of  majority class samples (Downsampled) can yield the best performance, since it preserves the original sample distribution for the minority class. We use SMOTE~\cite{chawla2002smote} for oversampling around the existing minority class distributions. While SMOTE (also bootstrapping) can be more effective when there are insufficient minority class samples, such oversampling (or re-sampling) can change the distributions with synthetic sample  instances (or modified sample importance) and the decision decision boundaries for good or bad. 

\begin{table}[ht]
\centering
\caption{Latency overhead multipliers (1$\times$ for imbalanced training) across different imbalanced training techniques, with simulated imbalanced ratio of 10.0 }
\label{tab:imbalance_overheads}
\resizebox{\columnwidth}{!}{
\begin{tabular}{c | c | c |c}
\toprule
 \textbf{Balanced} & \textbf{Downsampled} & \textbf{Bootstrapped} & \textbf{Oversampled} \\
\midrule
$1.30\times$ & $1.41\times$ & $2.46\times$ & $3.07\times$ \\
\bottomrule
\end{tabular}}
\end{table}

Table~\ref{tab:imbalance_overheads} summarizes the training overheads (latency overhead normalized to that of Standard) of various imbalanced training techniques, for a fixed imbalanced ratio of 10.0. The actual overheads can vary and may increase with larger tables and higher imbalanced ratio -- however, one can expect a similar trend where Balanced and Downsampled are relatively much cheaper than the others. However, given that training is such a small portion ($\le$ 2\%) of the end-to-end proxy model optimization process, we can afford to try more expensive techniques, like Oversampling, especially when there are too few minority class samples (e.g., $\le$ 100 sample instances).

\subsection{Role of Embedding Quality}
\label{sec:embedding_quality}

\begin{figure*}[!th] 
    \centering
    \includegraphics[width=\textwidth,height=0.28\textwidth,trim=0 0  0 1cm, clip]{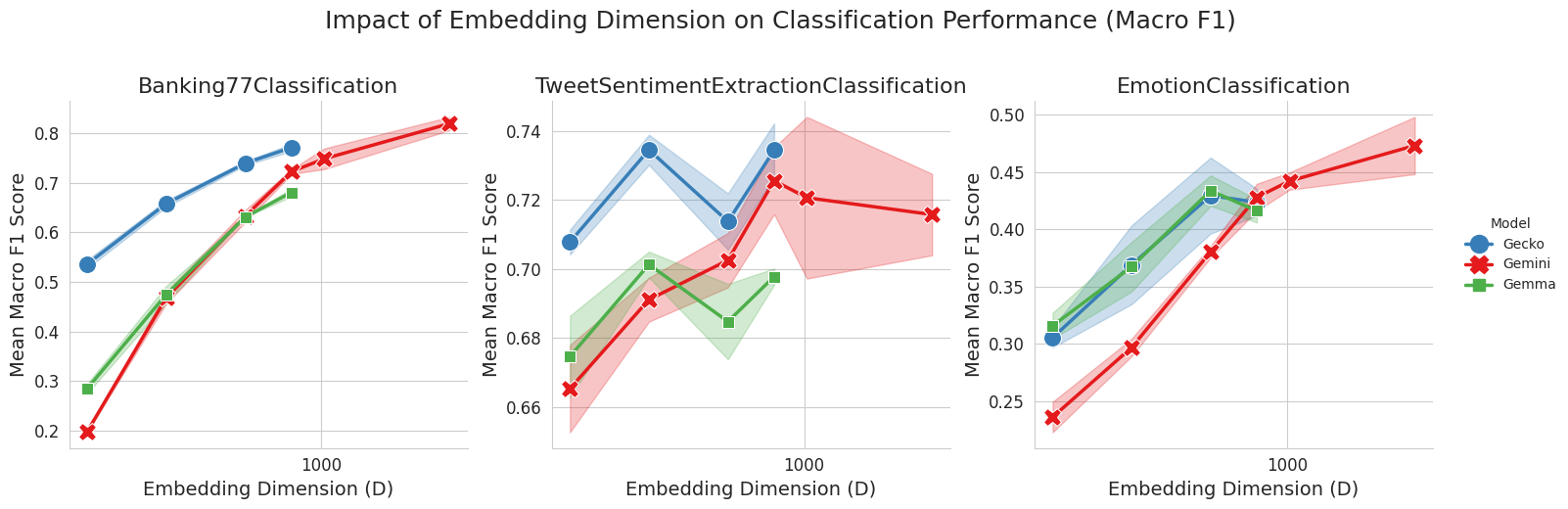}
    \caption{Impact of embedding model and dimensionality on proxy model classification performance. Note that Gecko (\textit{text-embedding-005}) and Gemma (\textit{embeddinggemma-300m}~\cite{Google_EmbeddingGemma300M}) support embedding dimension size up to 768, and Gemini (\textit{gemini-embedding-001}) supports up to 3072.}
    \label{fig:embedding_quality}
\end{figure*}

The selection of the input embedding model and the embedding vector dimension plays a crucial role in the success of the proxy approximation.
In Figure~\ref{fig:embedding_quality} we demonstrate how different embedding models of varying complexity and cost affect the classification performance of the trained proxy models.
The older, 768-dimensional Gecko (\textit{text-embedding-005}) model provides consistently high-quality text embeddings across most benchmarks (\yeounoh{see other benchmark results where we use Gecko model for proxy model training}). Notably, it outperforms the more expensive, larger 3072-dimensional Gemini (gemini-embedding-001) embeddings on the Tweet Sentiment extraction benchmark, suggesting that raw dimensionality does not always translate to superior quality for all tasks.
The open-source Gemma (\textit{embeddinggemma-300m}) model is the latest and most affordable option. However, proxy models trained using Gemma embeddings consistently exhibit the lowest performance, confirming that cost-savings at the embedding layer may translate into an unacceptable loss of accuracy for the downstream proxy classification.

\begin{table}[ht]
\centering
\caption{Embedding generation cost comparison for Tweet Sentiment test split (3534 rows)--this should increases linearly with the number of rows, since we are processing a fixed number of rows (20) per generation request.}
\label{tab:embedding_cost}
\begin{tabular}{
  l |  c | 
  >{\centering\arraybackslash}m{2.5cm}|
  c | c 
}
\toprule
\textbf{Model} & $\mathbf{D_{\text{max}}}$ & \textbf{Latency (sec)} & \textbf{Cost$^1$ (\$) } & \textbf{Size (MB)} \\
\midrule
Gemma   & 768  & 1x  & 1x  & 10.35 \\
Gecko   & 768  & 1.81x & 18.1x & 10.35 \\
Gemini  & 3072 & 4.18x & 27.6x  & 41.41 \\
\bottomrule
\multicolumn{5}{@{} p{\columnwidth} @{}}{
    \vspace{0.5ex}
    \footnotesize \raggedright
    $^1$ Cost is computed based on Gemini embedding model pricing~\cite{GoogleCloud_GenAIPricing} for Gecko and Gemini, and the cost of running a high CPU VM (8 vCPU, 16 GB RAM) for Gemma.
}
\end{tabular}
\end{table}

All three select embedding models use Matryoshka Representation Learning (MRL)~\cite{shaham2023matryoshka} to generate embeddings with prefixes working as a summary or simplified versions of the same information. 
While the models can generate reduced dimension embeddings with the same information, the compression results in the lower embedding quality -- observed across the models. In Table~\ref{tab:embedding_cost}, we compare the embedding generation costs of Gecko, Gemma and Gemini at their supported maximum embedding dimensions. Gemma is the fastest and also the most cost-efficient option. Gecko can be as effective as Gemini, \yeounoh{as shown in Figure~\ref{fig:embedding_quality}}, but at a much lower cost than Gemini.
Gemma's latency overhead and cost are used as references for normalization. The latency gap between Gemma and other models can be much larger on GPU, but it will also shrink the cost gap.
Note that Gemini embeddings at $D=3072$ are also much larger, \yeounoh{which leads to higher storage cost and can be non-trivial for large tables.}

\begin{figure*}[!th] 
    \centering
    \includegraphics[width=0.95\textwidth, height=0.45\textwidth]{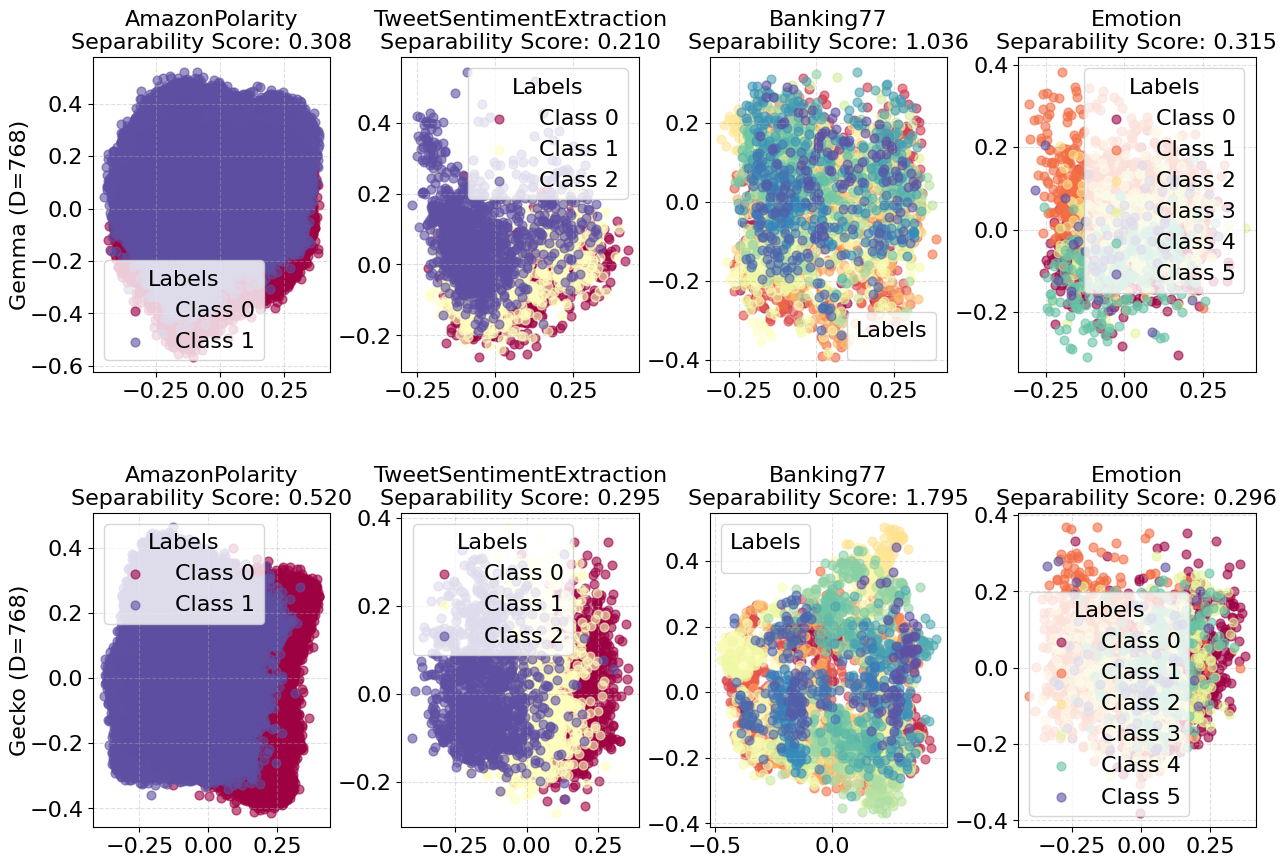}
    \caption{Embedding distinctiveness illustrated by PCA visualization (X-axis is PC1, Y-axis is PC2) and separability scores across various benchmark datasets with Gemma and Gecko embedding models, dimension $D=768$}
    \label{fig:embedding_pca}
\end{figure*}

Figure~\ref{fig:embedding_pca} presents a Principal Component Analysis (PCA) visualization comparing the distinctiveness of Gemma and Gecko embeddings ($D=768$) across four different classification benchmark datasets. 
\yeounoh{We perform PCA analysis of the each embedding vector to select two most informational principal components (dimensions) of the vector and them, as X and Y axes, to visualize the embeddings.}
Each plot is annotated with separability score \yeounoh{defined as a ratio between average inter class distance and average intra class variance.} Separability score measures how each cluster is compactly defined and isolated. The higher the separability score is the easier it is to separate and classify different classes. 
The overall takeaway is that Gecko is a better embedding model and reuslts in more distinctive clusters that are well separated, compared to those by Gemma. This explains why Gemma often results in inferior proxy approximation performances. Emotion classification dataset contains 6 different label clusters that are dispersed with significant overlap (low separability score). It is one of the harder benchmark dataset for our proxy approximation.

\subsection{Online VS. Offline Training and Model Selection}
\label{sec:learning_tuning}

\begin{table}[ht]
    \centering
    \caption{Normalized training latency to the LR training latency (1x) and tuned model performance on Tweet Sentiment dataset. }
    \label{tab:model_performance_tuning}
    \begin{tabular}{l|c|c|c|c}
        \toprule
        \textbf{Metric} & \textbf{LR} & \textbf{SVM} & \textbf{RF} & \textbf{XGB} \\
        \midrule
        F1 Score (default) & 0.867 & 0.861 & 0.841 & 0.858 \\
        F1 Score (tuned) & 0.867 & 0.861 & 0.836 & 0.848 \\
        \midrule
        Training Latency & 1x & 49.84x & 1.42x & 3.18x \\
        \bottomrule
    \end{tabular}
\end{table}

Table~\ref{tab:model_performance_tuning} summarizes the F1-scores  achieved with \textit{default} parameters versus  \textit{tuned} parameters via offline grid search, evaluated on the Tweet Sentiment dataset with four common classification algorithms: Logistic Regression (LR), Support Vector Machine (SVM), Random Forest (RF), and XGBoost (XGB). To our surprise, the benefit of hyperparameter tuning, while expensive and time-consuming offline, resulted in negligible or even slightly negative changes to the F1 score. 
For instance, the LR model retained an F1 score of 0.867 after tuning, showing no improvement. This lack of significant improvement was also observed across several other classification benchmark datasets, suggesting that the underlying embedding vectors are highly effective regardless of marginal model complexity. This finding eliminates the primary rationale for performing iterative, resource-intensive hyperparameter optimization during our online training process.

Table~\ref{tab:model_performance_tuning} also summarizes the normalized training latency overhead, \yeounoh{normalized to the LR training latency (1x)}, for different classification models using the default parameters. 
When comparing the normalized training latency LR stands out dramatically. LR represents the baseline (1x), while the next fastest model (RF) introduces a 1.42x overhead. SVM introduce a massive 49.84x overhead, and XGB requires 3.18x the time.
With LR providing the best balance of performance and speed (it is the fastest model and tied for the highest F1 score), we select Logistic Regression (LR) as the canonical online proxy model classification algorithm.

\subsection{Proxy Model Robustness Across Relational Filter Predicates}
In this section, we focus on the robustness of the proxy model when combined with relational filter predicates (e.g., segmenting data by age group, like AI.IF("review sentiment is positive") for age group A vs. B. 
In online training, we are training a proxy for each ad hoc query. In offline training, we pre-train a model for known query patterns and would like to re-use it for different data slices. It would be hard to pre-train a proxy model for all possible data cubes/slices.
It is crucial to determine whether a general proxy model, trained on a sample of the entire data population, remains effective when applied to specific data slices~\cite{chung2019automated,Chung2019SliceFA}.

\begin{table}[htbp]
    \centering
    \caption{Proxy and LLM performance (F1-Score) comparison across data slices of the California Housing dataset. This comparison highlights the difference between training on a global sample versus training specifically on sliced data.}
    \label{tab:combined_slice_analysis}

    \begin{subtable}[b]{0.4\textwidth}
        \centering
        \caption{Proxy model trained on global sample data (*).}
        \label{tab:f1_slice_analysis_a}
        \resizebox{\columnwidth}{!}{
        \begin{tabular}{l|c|c|c}
            \toprule
            \textbf{Slice ID} & \textbf{Proxy F1} & \textbf{LLM F1} & \textbf{Relative Accuracy} \\ 
            \midrule
            * & 0.945 & 0.908 & 1.040 \\
            \midrule
            0 & 0.977 & 0.993 & 0.984 \\
            1 & \textbf{0.617} & 0.334 & 1.847 \\
            2 & 0.980 & 0.992 & 0.988 \\
            3 & \textbf{0.244} & 0.100 & 2.440 \\
            4 & 0.977 & 0.998 & 0.979 \\
            5 & \textbf{0.841} & 0.575 & 1.463 \\
            6 & 0.960 & 0.977 & 0.983 \\
            7 & \textbf{0.596} & 0.355 & 1.679 \\
            \bottomrule
        \end{tabular}
        }
    \end{subtable}
    \hfill 
    \begin{subtable}[b]{0.4\textwidth}
        \centering
        \caption{Proxy model trained on slice-specific samples.}
        \label{tab:f1_slice_analysis_b}
        \resizebox{\columnwidth}{!}{
        \begin{tabular}{l|c|c|c}
            \toprule
            \textbf{Slice ID} & \textbf{Proxy F1} & \textbf{LLM F1} & \textbf{Relative Accuracy} \\ 
            \midrule
            0 & 0.972 & 0.993 & 0.978 \\
            1 & \textbf{0.296} & 0.334 & 0.886 \\
            2 & 0.991 & 0.992 & 0.998 \\
            3 & \textbf{0.091} & 0.100 & 0.910 \\
            4 & 0.977 & 0.998 & 0.979 \\
            5 & \textbf{0.508} & 0.575 & 0.883 \\
            6 & 0.960 & 0.977 & 0.983 \\
            7 & \textbf{0.292} & 0.355 & 0.822 \\
            \bottomrule
        \end{tabular}
        }
    \end{subtable}
\end{table}

Table~\ref{tab:combined_slice_analysis}(a) illustrates how the general proxy model trained on the entire California Housing dataset performs on eight mutually exclusive data slices, defined by a combination of three attributes: \textit{median\_house\_value}, \textit{median\_income}, \textit{latitude}.
\yeounoh{The (*) line shows the F1-score for the whole dataset, without any relational filter predicates.}
When LLM fails to produce accurate predictions (slices 1, 3, 5, 7), Proxy actually performs better. In fact, the relative accuracy close to 1.0 and for some, including the slices where LLM struggles, relative accuracy is much greater than 1.0.

\yeounoh{
Table~\ref{tab:combined_slice_analysis}(b) reports the performance comparison with proxy models trained on the samples taken from each target data slice.
Interestingly, training a specialized proxy model for a given slice did not yield meaningful performance improvements and rather performed worse (slices 1, 3, 5, 7) due to insufficient and/or imbalanced training data.}
While the robustness of a general proxy model trained for the entire table can be a data-dependent property, our findings suggest that a general proxy model still exhibit high relative accuracy and it can still perform reliably on most of the data slices, provided that the LLM is able to accurately label the slice sample data (when LLM accuracy is low, so is the proxy).

\section{Discussion and Lessons for Future Research}
\label{sec:discussions}

\subsection{Optimize Using Diverse Array of Models}
\label{sec:adaptive_ranker_selection}
Future query engines featuring semantic operators should leverage a diverse array of specialized proxy models for semantic operators, significantly enhancing the capability beyond simple classifiers and a single LLM baseline. This array includes simple, cost-effective alternatives such as logistic regression models and fine-tuned and/or small LLMs. In addition, for ranking, cross-attention ranking APIs (a.k.a. re-rankers) are beneficial for many use cases. While model cascading, as applied in prior work (e.g., \cite{patel2024lotus,jo2024thalamusdb}) to substitute smaller LLMs, offers a precedent for optimization and accuracy guarantees, utilizing the full diverse array of ``cheap" models is a primary avenue for research.

\subsection{Proxies for other AI Query Operators}\label{sec:other_operators}
While our primary focus is on AI.IF (semantic filter) and AI.RANK (semantic ranking), the proxy strategy is extensible to other semantic operators.

\begin{table}[t!]
\centering
\caption{AI.CLASSIFY performance comparison: LLM vs. Proxy Approximation (Precision / Recall)}
\label{tab:ai_classify_results}
\begin{tabular}{@{}lcccc@{}}
\toprule
\textbf{Dataset} & \textbf{LLM (P / R)} & \multicolumn{3}{c}{\textbf{Approx (P / R) by Sample Size}} \\ \cmidrule(l){3-5} 
 & & \textbf{1,000} & \textbf{4,000} & \textbf{8,000} \\ \midrule
BBC News & 0.94 / 0.94 & 0.95 / 0.95 & --- & --- \\ \addlinespace
DBpedia & 0.98 / 0.98 & 0.83 / 0.65 & 0.94 / 0.95 & 0.96 / 0.96 \\ 
\bottomrule
\multicolumn{5}{@{} p{\columnwidth} @{}}{
    \vspace{0.5ex}
    \footnotesize \raggedright
    BBC News has 5 categories, DBpedia~\cite{thedevastator_dbpedia_2022} has 14 categories.
}
\end{tabular}
\vspace{-0.5cm}
\end{table}

\noindent\textbf{Semantic Classification:} AI.IF proxy model can be extended to AI.CLASSIFY for multi-class classification. For logistic regression or many other common classification models, this extension only requires labeling each sample instance with non-binary labels. \yeounoh{A significant challenge remains the ``Needle-in-a-Haystack" sampling problem; obtaining sufficient representative examples for every class-label is essential for accurate model training, particularly for rare or minority classes. Table~\ref{tab:ai_classify_results} illustrates that the proxy approach could work for multi-label classification, but it requires more samples (random sampling) as the number of class labels increases.}

\noindent\textbf{Semantic Join:} 
A semantic join typically involves expensive LLM calls to evaluate conditions across join keys. While a proxy can be trained to predict these outcomes and accelerate execution (e.g., train a single multi-label classifier or a binary classifier for each distinct key from the smaller join table), a na\"{i}ve application still requires $O(N \times M)$ inferences. This remains prohibitively expensive at scale; for instance, joining two tables of 10,000 rows would require 100 million inferences. A complete solution requires integrating proxies with effective pre-filtering~\cite{patel2024lotus}, such as vector similarity, to limit the candidate pairs before proxy-based evaluation. While optimizing join algorithms is outside the current scope of this work, our evaluation of the proxy approach for classification tasks provides the foundation for such future systems.
Beyond scalability, the high selectivity of semantic joins poses the ``Needle-in-a-Haystack'' challenge observed in semantic ranking, where extremely low densities of relevant matches makes effective samplign and training nearly impossible (e.g., SciFact in Table~\ref{tab:ir_benchmark_results}). Consequently, the efficacy of a proxy-based join will highly depend on specific workload characteristics, such as join selectivity and the semantic complexity of the join predicate, calling for more specialized algorithms in future work.


\section{Conclusion \& Future Work}
\label{sec:conclusion}
In this paper, we studied the feasibility of non-LLM based proxy models as a viable alternative to LLMs for semantic operators, for both \textit{BigQuery} (OLAP) and \textit{AlloyDB} HTAP database setups. 
Through detailed experiments we demonstrated that proxy models can provide 100x latency and cost improvements while achieving similar accuracy as their LLM-based counterparts.  This is possible by reducing the AI.IF and AI.RANK operators into binary classification, careful stratified sampling and effective embedding models.  The key insight is that cost-effective large-scale OLAP and ultra-low latency HTAP using semantic operators are possible. Future optimizers should use such low cost proxy models, falling back to  costly LLM invocations for specific hard use cases and only when necessary. When such optimizers emerge, this study should be revised to measure the effectiveness of systems that automate the use of mixes of models.
 



\bibliographystyle{ACM-Reference-Format}
\bibliography{main}


\end{document}